\documentclass[a4paper,11pt]{article}
\usepackage{jheppub}
\usepackage{bm}

\title{Improved lattice fermion action for heavy quarks}

\author[a]{Yong-Gwi Cho,}
\author[b,c]{Shoji Hashimoto,}
\author[d]{Andreas J\"uttner,}
\author[b,c]{Takashi Kaneko,}
\author[d,e]{Marina Marinkovic,}
\author[b]{Jun-Ichi Noaki}
\author[d]{and Justus Tobias Tsang}

\affiliation[a]{Graduate School of Pure and Applied Sciences, 
  University of Tsukuba, Tsukuba, Ibaraki 305-8571, Japan}
\affiliation[b]{High Energy Accelerator Research Organization (KEK),
  Tsukuba 305-0801, Japan}
\affiliation[c]{School of High Energy Accelerator Science, The
  Graduate University for Advanced Studies (Sokendai), 
  Tsukuba 305-0801, Japan}
\affiliation[d]{School of Physics and Astronomy, University of Southampton, Highfield, Southampton SO17 1BJ, UK}
\affiliation[e]{CERN, Physics Department, 1211 Geneva 23, Switzerland}

\emailAdd{cho@ccs.tsukuba.ac.jp}
\emailAdd{shoji.hashimoto@kek.jp}
\emailAdd{a.juttner@soton.ac.uk}
\emailAdd{takashi.kaneko@kek.jp}
\emailAdd{marina.marinkovic@cern.ch}
\emailAdd{noaki@post.kek.jp}
\emailAdd{j.t.tsang@soton.ac.uk}

\subheader{\footnotesize UTHEP-673, KEK-CP-321, CERN-PH-TH-2015-074}

\abstract{
  We develop an improved lattice action for heavy quarks based on Brillouin-type fermions, that have excellent energy-momentum
  dispersion relation.  The leading discretization errors of $O(a)$ and $O(a^2)$ are eliminated at tree-level. We carry out a scaling study of this improved Brillouin fermion action on quenched lattices by calculating the charmonium energy-momentum dispersion relation and hyperfine splitting.  We present a comparison to standard Wilson fermions and domain-wall fermions. 
  }

\newcommand{\bvec}[1]{\mbox{\boldmath $#1$}}
  
\begin{document}
 \maketitle
\flushbottom

\section{Introduction}
Charm and bottom quarks have substantially shorter Compton wave-lengths than the typical length scale of Quantum Chromodynamics (QCD), $1/\Lambda_{\mathrm{QCD}}$. This poses a problem for numerical simulations of QCD on the lattice. The resolution of the lattice, the lattice spacing $a$, is chosen such that $a$ is sufficiently smaller than $1/\Lambda_{\mathrm{QCD}}$ while the entire lattice size $L$ has to be much larger than the length scale of the inverse pion mass, the lightest particle in the system. For the lattices that can be generated with currently available computational resources, the charm quark mass $m_c$ is similar to $1/a$, and the bottom quark mass $m_b$ is even larger. 
This was the motivation for introducing the static or non-relativistic effective theories for heavy quarks, which allow for disentangling the relevant physical scales in these calculations. The clear scale separation helps in the control of the systematics, but the effective theory approaches require an increasing number of extra terms and tuning their associated parameters in order to achieve more precise calculations. As an alternative to the effective heavy quark theories, in this work we perform an extensive feasibility study of different relativistic approaches to the heavy quark physics from the lattice.

Treating heavy quarks on the lattice with the conventional relativistic formulation has the advantage that the calculation can be made more and more precise as smaller lattice spacings become available. 
Currently, the finest lattices have $1/a\simeq$ 4~GeV and the attempts are being made to raise it  to 5--6~GeV in the coming years. 
The use of the relativistic fermion formulations is therefore a promising option in the near future. 
For that to be really useful, it is essential to use improved fermion discretizations that allow to make precise predictions even when $m$ is not much smaller than $1/a$.
One successful example is the Highly Improved Staggered Quark (HISQ) formulation \cite{Follana:2006rc}, for which the staggered fermion formulation is improved by introducing higher dimensional operators, and the leading discretization error is of order $(am)^4$ for heavy quarks. This formulation has been applied for a number of calculations of phenomenologically important quantities, such as $D_{(s)}$ and $B_{(s)}$ meson decay constants and other form factors \cite{Follana:2007uv,Davies:2010ip,McNeile:2011ng,Na:2010uf,Na:2011mc}.

Among other relativistic actions, which do not involve the complication due to the fermion doubling of the staggered fermion formulations, the widely used formulations still contain the discretization effects of $O(a^2)$, which have to be eliminated to achieve a similar level of precision to that of the HISQ formulation. This can be done in a systematic way according to the recipe of the Symanzik improvement program \cite{Symanzik:1983dc,Symanzik:1983gh}, and some attempts were made in the past \cite{Sheikholeslami:1985ij,Eguchi:1983xr,Hamber:1983qa,Alford:1996nx} but they have not been used extensively except for the minimal one,  {\it i.e.} the $O(a)$-improved (or clover) action \cite{Sheikholeslami:1985ij}, mainly because the non-perturbative tuning of improvement parameters requires a lot of effort.

The goal of this work is to study the scaling of relativistic heavy-quark formulations in the quenched approximation, before dynamical configurations with similar parameters become available. In particular, we present the scaling study of heavy-heavy meson correlators, while the scaling of the heavy-strange systems will be presented in a future publication. 

In this paper, we mainly describe a study of the fermion formulation based on the improved covariant derivative and Laplacian operators \cite{Durr:2010ch}. We compare this Brillouin fermion formulation to more standard lattice fermions, such as the non-improved Wilson fermion formulation and the M\"obius domain Wall fermions (non-smeared and smeared)  \cite{Brower:2012vk}. In order to investigate the scaling towards the continuum limit, we generate lattice gauge ensembles in the range of $1/a$ = 2.0--5.6~GeV in the quenched approximation and perform the measurements of heavy-heavy correlators. 

Among many options explored in \cite{Durr:2010ch}, we consider a combination of the ``isotropic'' covariant derivative (iso) and ``Brillouin'' Laplacian (bri). This so-called Brillouin fermion is designed such that the violation of four-dimensional rotational symmetry is minimized. By such modification, it turned out that the energy-momentum dispersion relation of a massless fermion is much closer to the continuum one compared to that of a standard Wilson fermion \cite{Durr:2010ch}.  In the context of the Symanzik improvement, this is not obvious since the leading discretization error of $O(a)$ remains with this prescription. But, as far as the tree-level dispersion relation is concerned, the improvement seems to be achieved including higher orders of the lattice spacing $a$. Once the dispersion relation is improved, one can expect that interaction terms are also improved, since the form of fermion and gauge field interaction is highly constrained in the gauge theory. Namely, one simply replaces the tree-level derivative terms by the corresponding covariant derivatives by inserting the gauge links.

In this work, we consider a further improvement of the Brillouin-based fermion formulation according to the Symanzik improvement program. We design the lattice action such that the discretization effects of $O(a)$ and $O(a^2)$ are eliminated at the tree-level. With our choice we find that the continuum-like energy-momentum dispersion relation is satisfied very precisely for quark masses up to $am\sim 0.5$.

Another virtue of Brillouin fermions can be seen in its eigenvalue distribution in the complex plane. Unlike the standard Wilson fermion formulation, the Brillouin fermion has eigenvalues which lie very closely on the unit circle which the Ginsparg-Wilson relation \cite{Ginsparg:1981bj} requires. It suggests that this fermion formulation has an approximate chiral symmetry without explicitly constructing the overlap operator of \cite{Neuberger:1997fp,Neuberger:1998wv}. It also means that the Brillouin-Dirac operator is suitable as a kernel of the overlap operator and relatively small numerical effort is needed to build the overlap operator. We mention this possibility and its improvement beyond $O(a^2)$.

The mentioned properties of the Brillouin-type fermions are not guaranteed to be satisfied beyond tree-level, and a non-perturbative
study is needed to test the size of the scaling violations in the interacting case. In this work we explicitly check the scaling towards the continuum limit by taking some basic non-perturbative quantities, such as the heavy-meson dispersion relation and hyperfine splitting.

This paper is organized as follows. In Section~\ref{sec:formulation} we review the construction of the Brillouin-type fermion and study its improvement according to the Symanzik improvement program. At tree-level, we compare the energy-momentum dispersion relation and complex eigenvalue spectrum of the Dirac operator of various formulations.  The improved Brillouin fermion has a limitation on the values of quark mass  due to a violation of the reflection positivity property as discussed in Section~\ref{sec:heavy_problem}. Section~\ref{sec:scaling_study} describes a non-perturbative scaling study of the improved Brillouin fermion and its comparison to the standard Wilson fermion and domain-wall fermions. We then conclude in Section~\ref{sec:conclusion}.

\section{Definition and tree-level analysis}
\label{sec:formulation}

\subsection{Brillouin operators}
The Brillouin-type covariant derivative and Laplacian operators were introduced in \cite{Durr:2010ch}. 
(See also, \cite{Creutz:2010bm}, which introduced similar types of operators in a different context.)
We write the lattice Dirac operator as
\begin{eqnarray}
  S_{F} & = & 
  \sum_{n,m} \overline{\psi}_{n}D\left(n,m\right)\psi_{m}, 
  \\
  D(n,m) & = & 
  \sum_\mu \gamma_\mu\nabla_\mu(n,m)
  -\frac{a}{2}\triangle\left(n,m\right) + m_{0}\delta_{n,m}   \nonumber \\
  & &  
  \; \; \; \; \; \; \; \; \; \; \; \; \; \; \; \; \; \; \; \;\; \; \; \; \; \; \; \;\; \;\; \; \; \; \; \; \; \
  - \frac{c_{sw}}{2} \sum_{\mu<\nu} \sigma_{\mu\nu}F_{\mu\nu}\delta_{n,m},
  \label{eq:Dirac_op}
\end{eqnarray}
where $\nabla_{\mu}(n,m)$ and $\triangle(n,m)$ are the generalized covariant derivative term and Laplacian, respectively. The Sheikholeslami-Wohlert (or clover) term \cite{Sheikholeslami:1985ij} could also be introduced with a coefficient $c_{sw}$ when one introduces the field rotation for the $O(a)$-improvement, but we do not consider this possibility in this paper.

For the standard Wilson fermion, the derivative operators are
\begin{eqnarray}
  \label{eq:nabla_std}
  \nabla_\mu^{std}(n,m) & = 
  \frac{1}{2a} (\delta_{n+\hat{\mu},m}-\delta_{n-\hat{\mu},m}),
  \\
  \label{eq:triangle_std}
  \triangle^{std}(n,m) & = 
  \frac{1}{a^2} \sum_\mu
  (\delta_{n+\hat{\mu},m}-2\delta_{n,m}+\delta_{n-\hat{\mu},m})
\end{eqnarray}
at tree-level; the gauge interaction is introduced by promoting the hopping terms $\delta_{n\pm\hat{\mu},m}$ to a covariant derivative including a gauge link.
In momentum space, they are given as
\begin{eqnarray}
  \tilde{\nabla}_\mu^{std}(p) & = &
  \frac{i}{a} \sin(p_\mu a)
  = i\left( p_\mu - \frac{a^2}{6} p_\mu^3 + O(a^4) \right),
  \\
  \tilde{\Delta}^{std}(p) & = &
  \frac{2}{a^2} \sum_\mu \left(\cos(p_\mu a)-1\right)
  = - p^2 + O(a^4).
  \vspace{-5mm}
\end{eqnarray}
The leading discretization effects are ones from $\nabla_\mu^{std}$ of $O(a^2)$, as well as those of $a\triangle^{std}$, which is $O(a)$. We note that the $O(a^2)$ term of $\nabla_\mu^{std}$ violates rotational and Lorentz symmetry.

Among many options proposed in \cite{Durr:2010ch}, the choice of $\nabla^{iso}_\mu$ and $\triangle^{bri}$ leads to the most continuum-like dispersion relation.
Their explicit forms are
\begin{eqnarray}
  \label{eq:nabla_iso}
  \nabla^{iso}_{\mu}(n,m) & = &
  \rho_1 [\delta_{n+\hat{\mu},m}-\delta_{n-\hat{\mu},m}]
  + \rho_2 \sum_{\nu(\neq\mu)}[ \delta_{n+\hat{\mu}+\hat{\nu},m} -
                     \delta_{n-\hat{\mu}+\hat{\nu},m} ] 
  \nonumber \\
  & &
  + \rho_3 \sum_{\nu\neq\rho(\neq\mu)} [
      \delta_{n+\hat{\mu}+\hat{\nu}+\hat{\rho},m} -
      \delta_{n-\hat{\mu}+\hat{\nu}+\hat{\rho},m} ]
  \nonumber\\
  & &
  + \rho_4 \sum_{\nu\neq\rho\neq\sigma(\neq\mu)} [
      \delta_{n+\hat{\mu}+\hat{\nu}+\hat{\rho}+\hat{\sigma},m} -
      \delta_{n-\hat{\mu}+\hat{\nu}+\hat{\rho}+\hat{\sigma},m} ] , 
 \end{eqnarray}
 \begin{eqnarray}
  \label{eq:laplacian_bri}
  \triangle^{bri}(n,m) & = &    
  \lambda_0 \delta_{n,m} 
  + \lambda_1 \sum_\mu \delta_{n+\hat{\mu},m}
  + \lambda_2 \sum_{\mu\neq\nu} \delta_{n+\hat{\mu}+\hat{\nu},m} \nonumber \\
  & &
  + \lambda_3 \sum_{\mu\neq\nu\neq\rho} \delta_{n+\hat{\mu}+\hat{\nu}+\hat{\rho},m}
  + \lambda_4 \sum_{\mu\neq\nu\neq\rho\neq\sigma}
              \delta_{n+\hat{\mu}+\hat{\nu}+\hat{\rho}+\hat{\sigma},m}
\end{eqnarray}
with $(\rho_1,\rho_2,\rho_3,\rho_4)=\frac{1}{432}(64,16,4,1)$ and $(\lambda_0,\lambda_1,\lambda_2,\lambda_3,\lambda_4)= \frac{1}{128}(240,-8,-4,-2,-1)$.
The summations in (\ref{eq:nabla_iso}) and (\ref{eq:laplacian_bri}) run over positive and negative directions, $\mu,\nu,\rho,\sigma=\pm1,\pm2,\pm3,\pm4$ and all indices are different from one another, {\it i.e.} $\mu\neq\nu\neq\rho\neq\sigma$. Under this restriction, these operators connect neighboring lattice sites $m$ in a $3^4$ hypercube with $n$ in its center. By counting hops along the gauge links, they have up to four hops.

In order to make these operators gauge covariant, we have to insert gauge links for each hop. This should be done so that the rotational symmetry under the cubic group is respected. We average the shortest possible paths in the taxi-driver distance. For two-hop terms there are two paths; three-hop terms have six paths. The most complicated four-hop terms have 24 shortest paths to be averaged. For practical implementation of them, see Appendix~\ref{sec:hops}.

In momentum space (at tree level), they have the form
\begin{eqnarray}
  \label{eq:nabla_iso_mom}
  \tilde{\nabla}_\mu^{iso}(p) 
  & = & 
  \frac{i}{27a}\sin(p_\mu a)
  \prod_{\nu\neq\mu} \left[\cos(p_\nu a)+2\right]
  \nonumber\\
  & = &
  \frac{i}{27} p_\mu \left[
    \left(1-\frac{1}{6}(p_\mu a)^2\right)
    \prod_{\nu\neq\mu}\left(3-\frac{1}{2}(p_\nu a)^2\right)
    + O(a^4)
  \right]
  \nonumber\\
  & = &
  ip_\mu \left[ 1 - \frac{1}{6} (pa)^2 + O(a^4) \right]
 \end{eqnarray}
and  
 \begin{eqnarray}
  \label{eq:laplacian_bri_mom}
  \tilde{\triangle}^{bri}(p) & = &   
  \frac{4}{a^2} \left[
    \prod_\mu \cos^2\left(\frac{p_\mu a}{2}\right)-1 
  \right]
  \nonumber\\
  & = &
  - p^2 + O(a^4).
\end{eqnarray}
The derivative operator $\nabla_\mu^{iso}$ has $O(a^2)$ discretization effects which are invariant under rotation, thus the name of ``iso''.

\begin{figure}[tbp]
  \includegraphics[clip,scale=0.35]{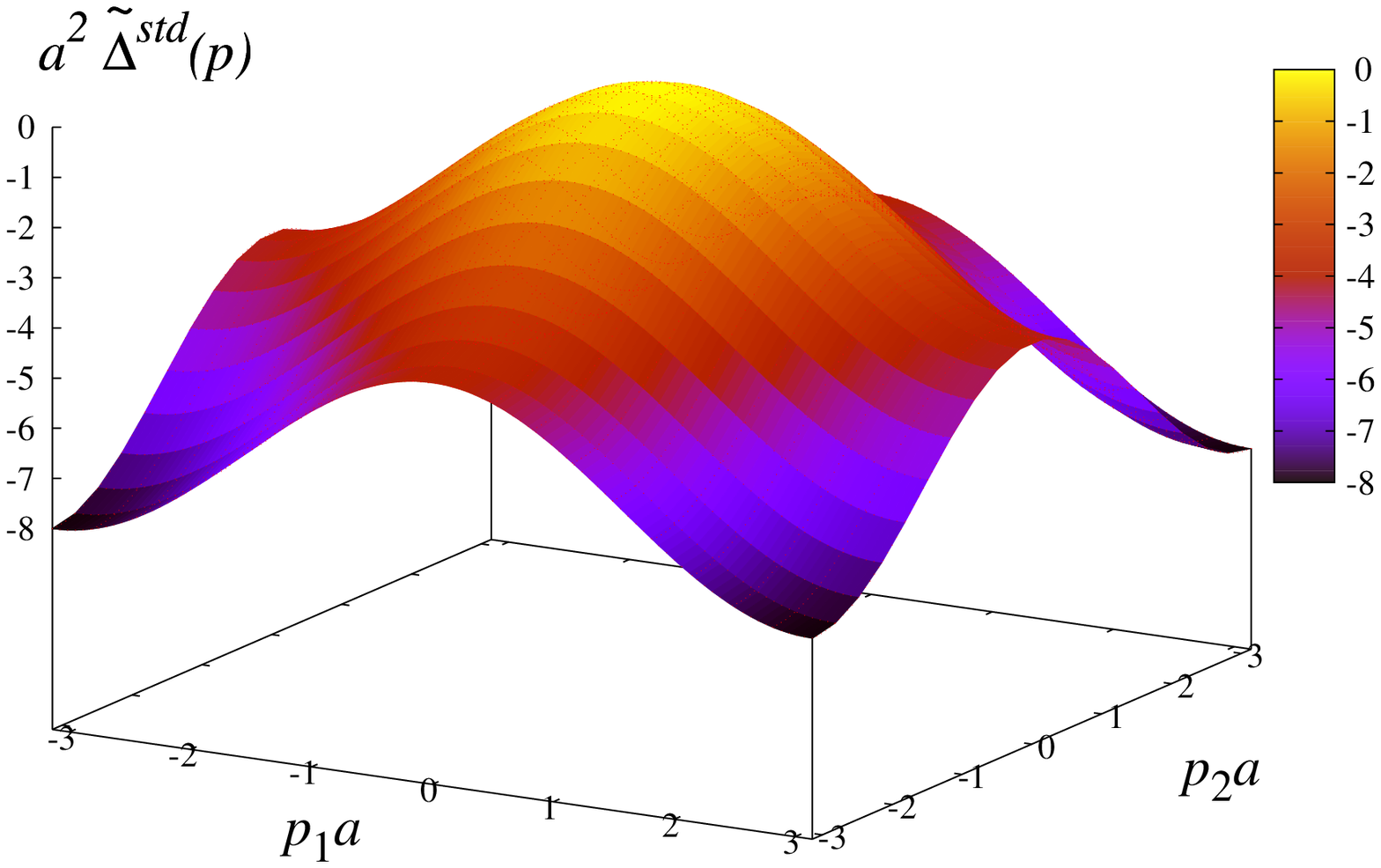}
  \includegraphics[clip,scale=0.35]{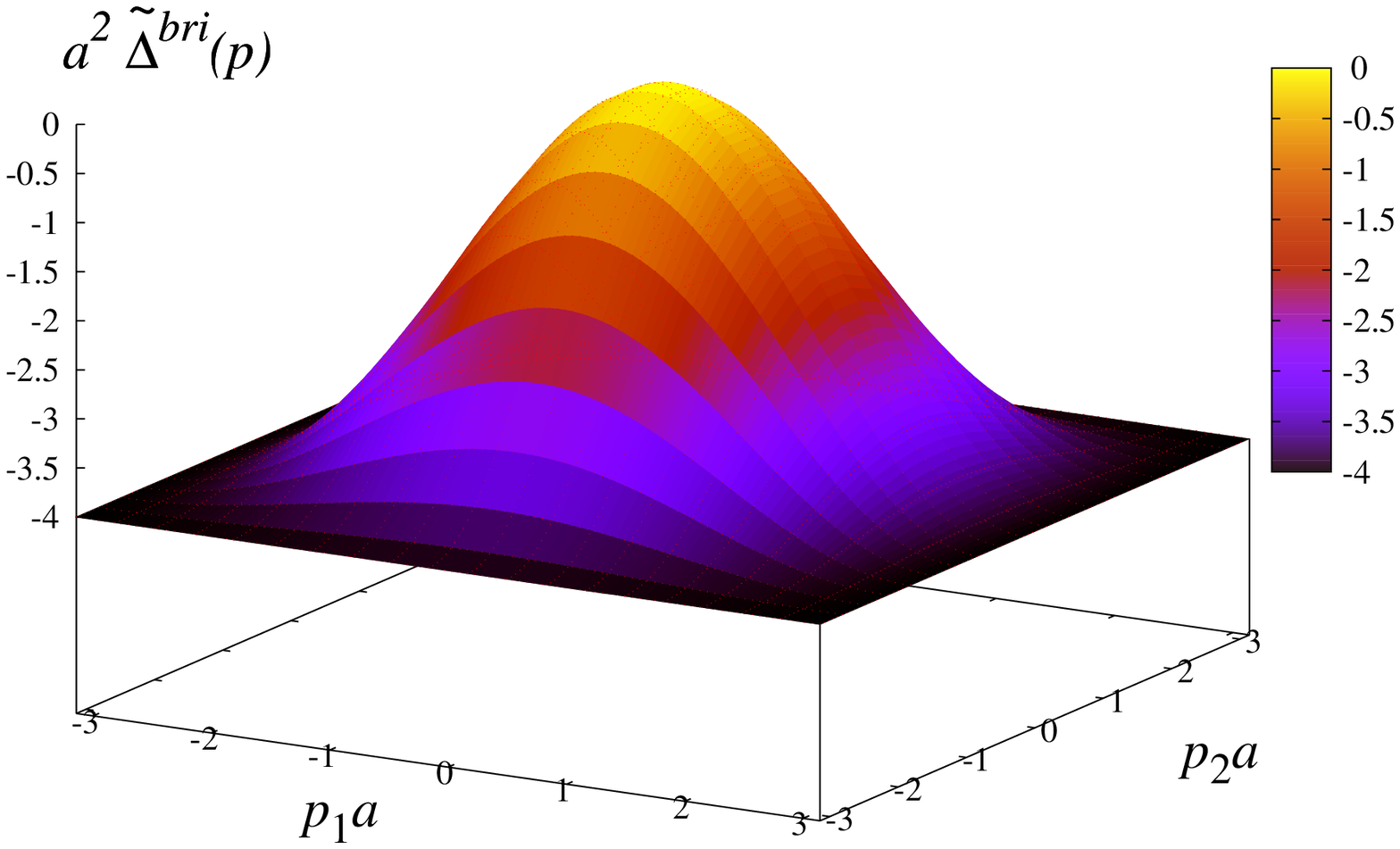}
  \caption{
    Laplacian operator $\tilde{\triangle}(p)$ shown in a
    two-dimensional momentum space $(ap_1,ap_2)$ (Other momentum
    components are assumed to be zero.)
    The standard $\tilde{\triangle}^{std}$ (left) and Brillouin
    $\tilde{\triangle}^{bri}$ (right) are shown.
  }
  \label{fig:laplacian}
\end{figure}

\begin{figure}[tbp]
  \centering
  \includegraphics[scale=0.4]{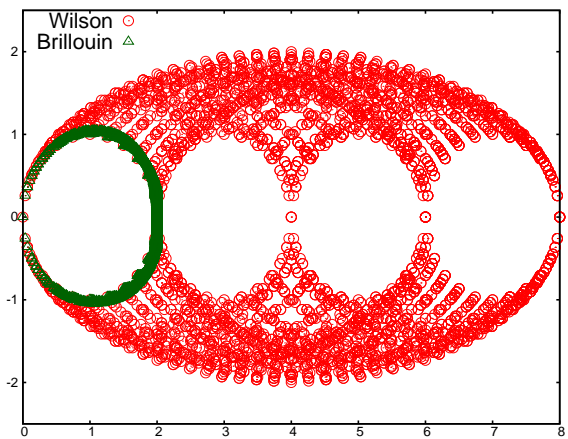}
  \caption{
    Eigenvalues of the Dirac operator on a complex plane.
    They are calculated on a free gauge field background for the Wilson fermion (red circles) and the Brillouin fermion (filled green triangles).
    \label{fig:Eigenvalue-wil_vs_bri}
  }
\end{figure}

The Brillouin-type Laplacian (\ref{eq:laplacian_bri}) has the interesting structure that the doublers on the edges of the Brillouin zone have the same mass. Indeed, for (non-zero) momenta  $ap_\mu=(\pm\pi,\pm\pi,\pm\pi,\pm\pi)$, the form in the momentum space (\ref{eq:laplacian_bri_mom}) implies that the induced mass is always $2/a$. Figure~\ref{fig:laplacian} shows $\tilde{\triangle}^{std}$ and $\tilde{\triangle}^{bri}$ in two-dimensional space. It apparently shows that the Brillouin-type Laplacian shows a flat tail at the edge of the Brillouin zone. This can also be seen from the eigenvalue spectrum of the Dirac operator. Figure~\ref{fig:Eigenvalue-wil_vs_bri} shows the eigenvalues of the Dirac operator $D(n,m)$ for the Wilson and the Brillouin operators calculated on a free gauge field background. The eigenvalues of the Wilson-Dirac operator plotted on a complex plane show five branches on the real axis, corresponding to the doublers of masses $0$, $2/a$, $4/a$, $6/a$ and $8/a$. For the Brillouin operator the doublers are all degenerate at $2/a$. Apart from the real axis, the eigenvalues roughly lie on a single orbit, very similar to those of the overlap-Dirac operator. It suggests that the operator is close to the overlap operator and the Ginsparg-Wilson relation is satisfied with good accuracy at least in the free field case. Among similar lattice fermion formulations which involve hopping terms within the $3^4$ hypercube \cite{Bietenholz:1999km,Bietenholz:2000iy,Gattringer:2000js,Gattringer:2000qu}, the Brillouin fermion is advantageous for both the continuum-like dispersion relation and the eigenvalue spectrum that approximately respects the Ginsparg-Wilson relation.

\subsection{Tree-level dispersion relation}
One useful measures of the discretization effect is the energy-momentum dispersion relation. It is defined through a pole of the fermion propagator, and takes the form $E=\sqrt{m^2+\bm{p}^2}$ in the continuum theory. For the lattice Dirac operator (\ref{eq:Dirac_op}), the pole is a solution of 
\begin{equation}
  \label{eq:pole}
  \left(\frac{1}{2}\tilde{\triangle}(p)-ma\right)^2
  - \sum_\mu \left(\tilde{\nabla}_{\mu}(p)\right)^2 = 0
\end{equation}
for specific forms of $\nabla_\mu$ and $\triangle$. The poles exist in the Minkowski region that is identified by assigning the ``energy'' $E$ as $p_4=iE$. There are more than one poles due to the doublers which are heavier than the physical mode by $O(1/a)$. In the following we only show the dispersion relation for the physically relevant pole unless otherwise stated.

\begin{figure}[tbp]
  \includegraphics[scale=0.33]{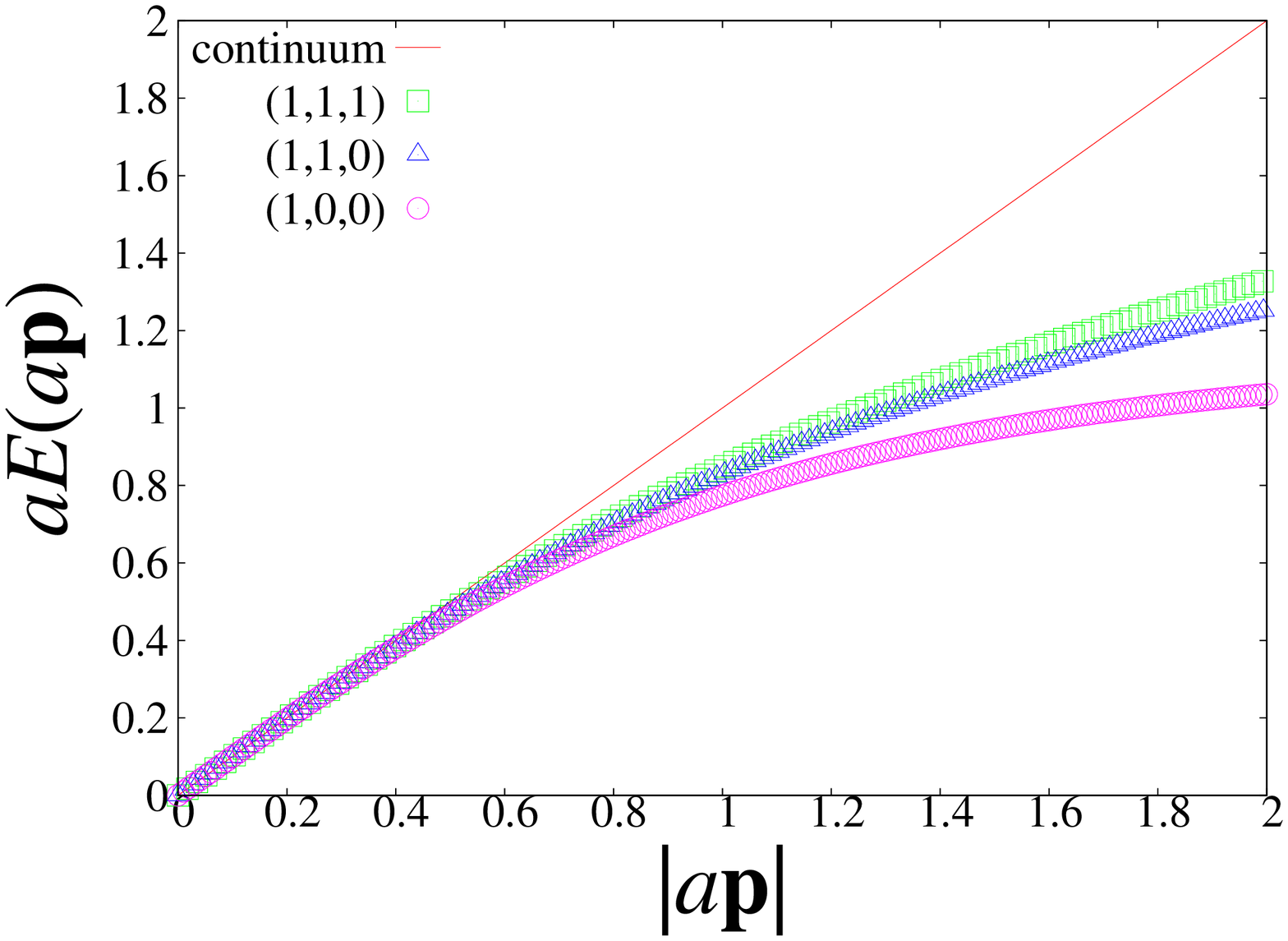}
  \includegraphics[scale=0.33]{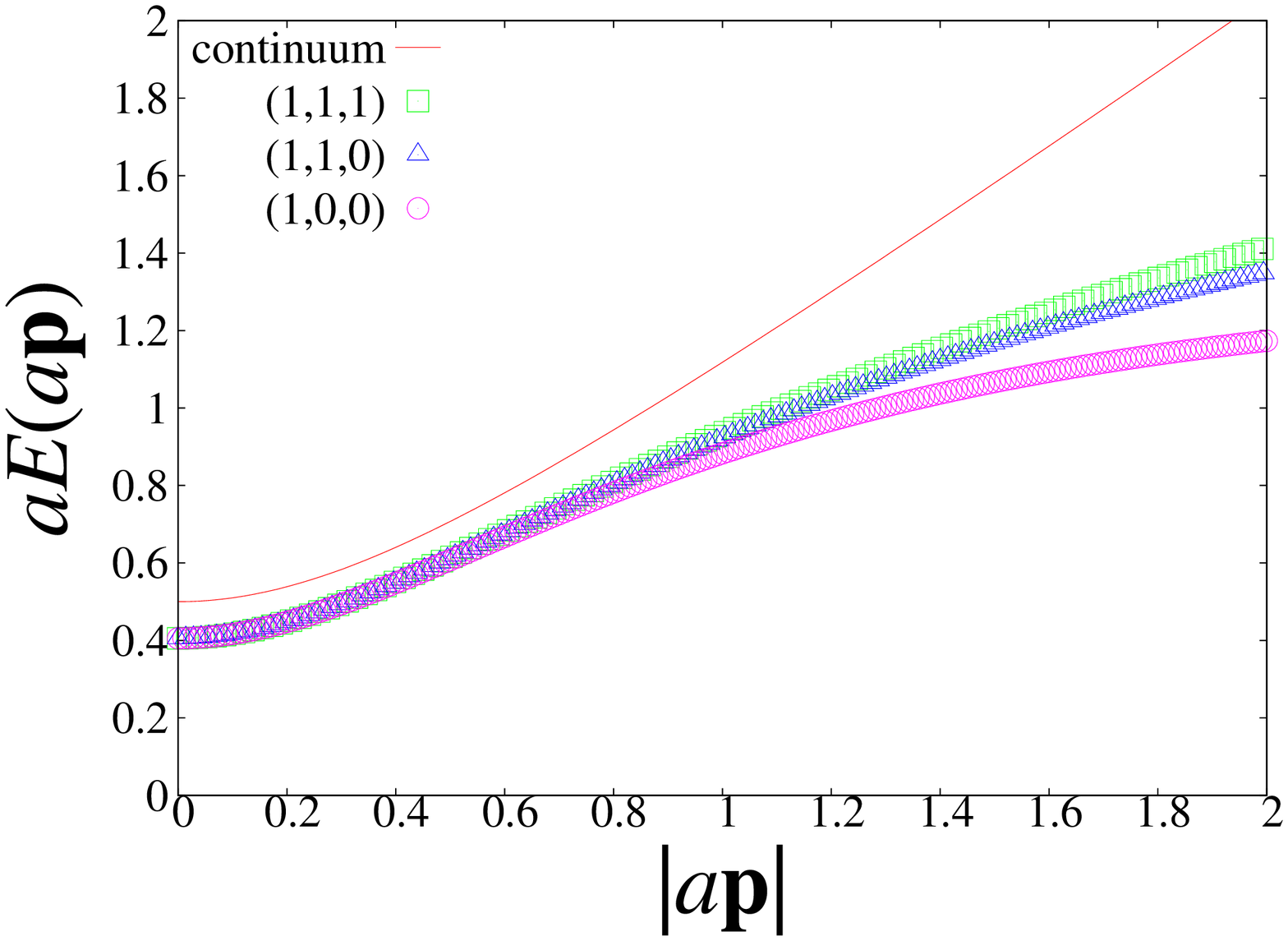}
  \caption{
    Energy-momentum dispersion relation for Wilson fermions.
    Two plots show the relation in the massless limit $am=0$ (left)
    and a massive  case of $am=0.5$ (right).
    Horizontal axis is the spatial momentum
    $|a\bm{p}|\equiv\sqrt{(a\bm{p})^2}$ in three directions parallel
    to (1,0,0), (1,1,0) and (1,1,1).
    Corresponding continuum relation $E=\sqrt{m^2+\bm{p}^2}$ is shown
    by a solid line.
}
  \label{fig:dispersion_Wilson}
\end{figure}

\begin{figure}[tbp]
  \includegraphics[scale=0.33]{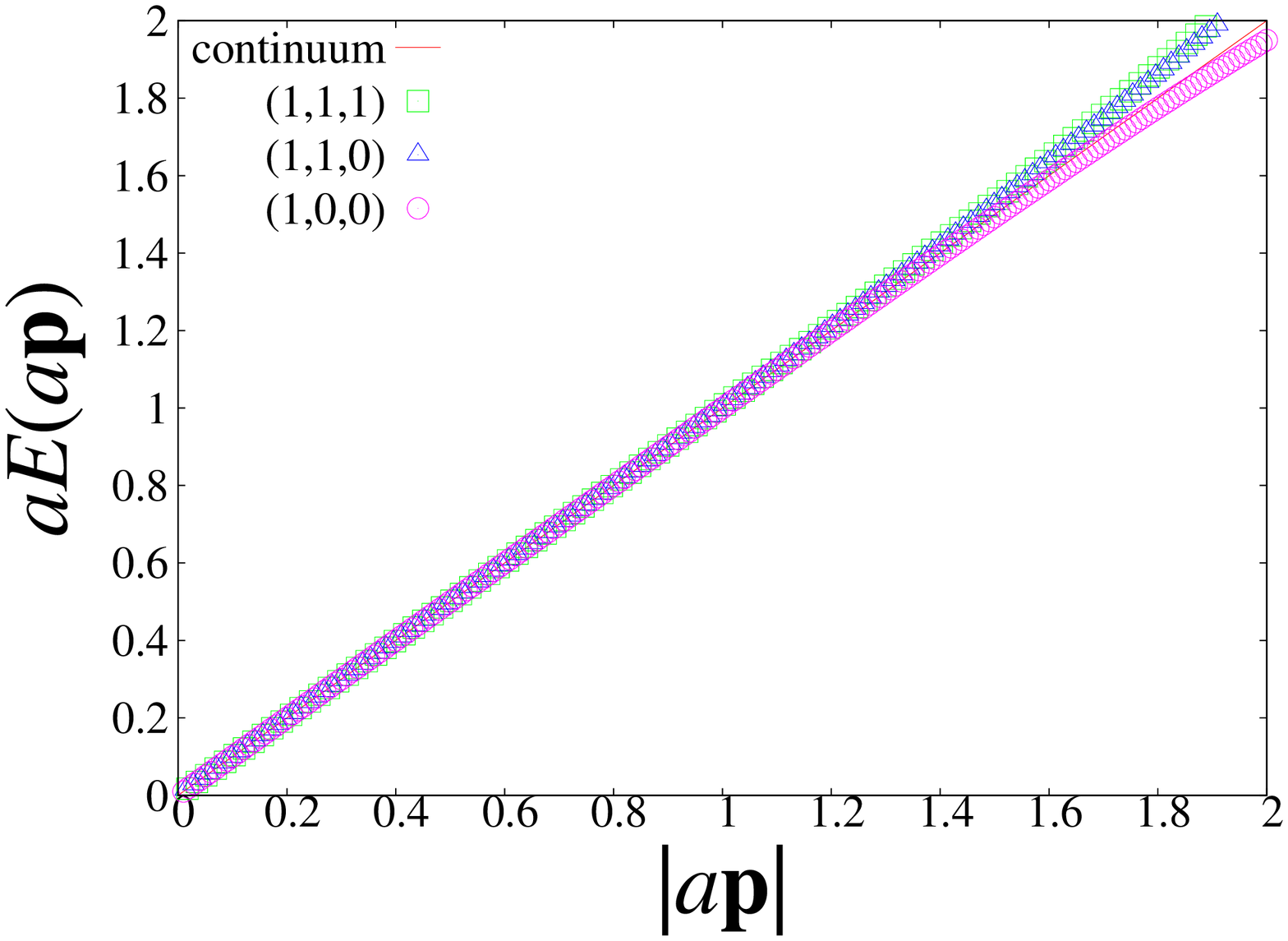}
  \includegraphics[scale=0.33]{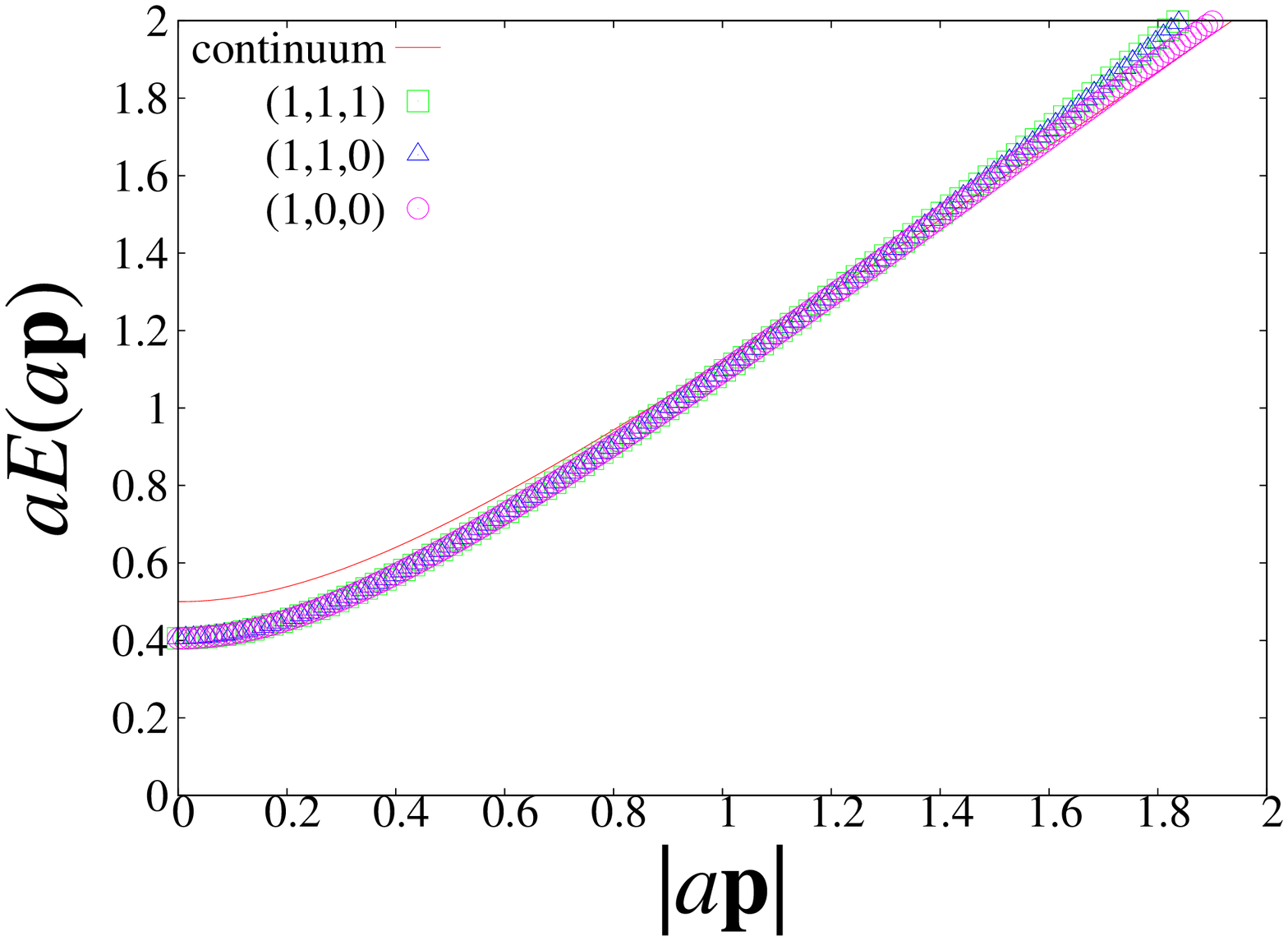}
  \caption{Same as Figure~\ref{fig:dispersion_Wilson}, but for the
    Brillouin fermion.}
  \label{fig:dispersion_Brillouin}
\end{figure}

The tree-level dispersion relations are shown in Figure~\ref{fig:dispersion_Wilson} and \ref{fig:dispersion_Brillouin} for Wilson and Brillouin fermions, respectively. As we have an application to heavy fermions in mind, we show the results for the massive case $am=0.5$ (right panel) as well as those in the massless limit (left). Lattice momenta are taken in three directions parallel to (1,0,0), (1,1,0) and (1,1,1), in order to see discretization effects which may violate rotational symmetry. The continuum relation $E=\sqrt{m^2+\bm{p}^2}$ is shown by a solid line, as well.

In the massless limit (left panels), the discretization effect is quite significant for Wilson fermions beyond $|a\bm{p}|\gtrsim 0.5$, while the dispersion relation for the Brillouin fermion closely follows that of the continuum theory up to $|a\bm{p}|\simeq1.5$. For the massive case, $am=0.5$ (right panels), the deviation from the continuum curve is sizable for both Wilson fermions and Brillouin fermions already at $|a\bm{p}|=0$.  
If we shift the overall energy such that the dispersion relation agrees with the continuum one as adopted in the non-relativistic effective theory approaches,  the deviation would become visible above $|a\bm{p}|\sim$ 0.6. Still, the dispersion relation of the massive Brillouin fermion closely follows that of the continuum compared to the Wilson fermion.
The closeness to the continuum theory is quantified by Taylor-expanding the dispersion relation. Up to fifth order of $a$, we obtain

\begin{eqnarray}
  (aE)^2(a\bm{p},am) & = &
  \left[
    (am)^2-(am)^3+\frac{11}{12}(am)^4-\frac{5}{6}(am)^5
  \right]
  \nonumber\\
  & &
  + \left[
    1-\frac{2}{3}(am)^2+\frac{7}{6}(am)^3
  \right] (a\bm{p})^2
  \nonumber\\
  & &
  + \left[
    -\frac{2}{3}+\frac{am}{2}
  \right]
  \left(\sum_{i<j} a^4 p_i^2p_j^2 +\sum_i(ap_i)^4\right)
\end{eqnarray}
for Wilson fermions. The first line corresponds to an expansion of the exact relation $aE=\ln(1+am)$, which contains $O(a)$ discretization effects. The third line represents the terms that violate rotational symmetry. On the other hand, the expansion for the Brillouin fermion gives

\begin{eqnarray}
  (aE)^2(a\bm{p},am) & = &
  \left[
    (am)^2-(am)^3+\frac{11}{12}(am)^4-\frac{5}{6}(am)^5
  \right]
  \nonumber\\
  & &
  + \left[
    1+\frac{1}{12}(am)^3
  \right] (a\bm{p})^2
  \nonumber\\
  & &
  + \left[ \frac{ma}{12} \right]
  \left(\sum_{i<j} a^4 p_i^2p_j^2 +\sum_i(ap_i)^4\right).
\end{eqnarray}
There is no difference in the first term, since $\tilde{\nabla}_{\mu}^{iso}(a\bm{p}=0)=\tilde{\nabla}_{\mu}^{std}(a\bm{p}=0)$ and $\tilde{\triangle}^{bri}(a\bm{p}=0)=\tilde{\triangle}^{std}(a\bm{p}=0)$. For finite momenta the Brillouin fermion is improved: the coefficient of $(a\bm{p})^2$ does not have terms of $O((am)^2)$, and the rotational symmetry violating term is suppressed by another order of $a$. The second property follows from the fact that $\tilde{\nabla}_\mu^{iso}(p)$  has only an isotropic error at $O(a^2)$.

\subsection{D34 action}
One may wonder whether the improvement obtained with the Brillouin fermion might also be achieved by more traditional improved actions which include next-to-nearest neighbor interactions, such as those of  Eguchi and Kawamoto \cite{Eguchi:1983xr} or  Hamber and Wu \cite{Hamber:1983qa}. We call them the D34 action following the terminology of \cite{Alford:1996nx}. The Dirac operator is given as

\begin{equation}
  D_{D34} = \sum_\mu \gamma_\mu\nabla_\mu^{std}
  \left(1-\frac{1}{6}a^{2}\triangle_\mu^{std}\right) + 
  c_{D34} \sum_\mu  a^3 \left(\triangle_\mu^{std}\right)^2,
\end{equation}
where $c_{D34}$ is a free parameter. $\nabla_{\mu}^{std}$ is already defined in (\ref{eq:nabla_std}) and $\triangle_\mu^{std}$ is given by 

\begin{equation}
  \triangle^{std}_\mu (n,m) = \frac{1}{a^2} \left(
    \delta_{m,n+a\hat{\mu}}+\delta_{m,n-a\hat{\mu}}-2\delta_{m,n}
  \right).
\end{equation}

Note that the D34 action is defined without the fermion field rotation. Following the steps of calculating the energy-momentum dispersion,  we obtain an expansion for small $am$ and $a\bm{p}$ up to $a^5$ as

\begin{eqnarray}
  (aE)^2(a\bm{p},am) & = &
  \left[ (am)^2 + 2c_{D34}(am)^5 \right]
  \nonumber\\
  & &
  + \left[ 1 + 4c_{D34}(am)^3 \right] (a\bm{p})^2
  \nonumber\\
  & &
  + \left[ 4c_{D34} am \right]
  \left(\sum_{i<j} a^4 p_i^2p_j^2 + \sum_i(ap_i)^4 \right).
\end{eqnarray}
Therefore, it is improved so that there is no $O(a)$ and $O(a^2)$ term, as designed, while the Brillouin fermion contains errors of $O(a)$ and $O(a^2)$ in the term of vanishing momenta (the first line). In this sense, the D34 is even better.

\begin{figure}[tbp]
  \includegraphics[scale=0.33]{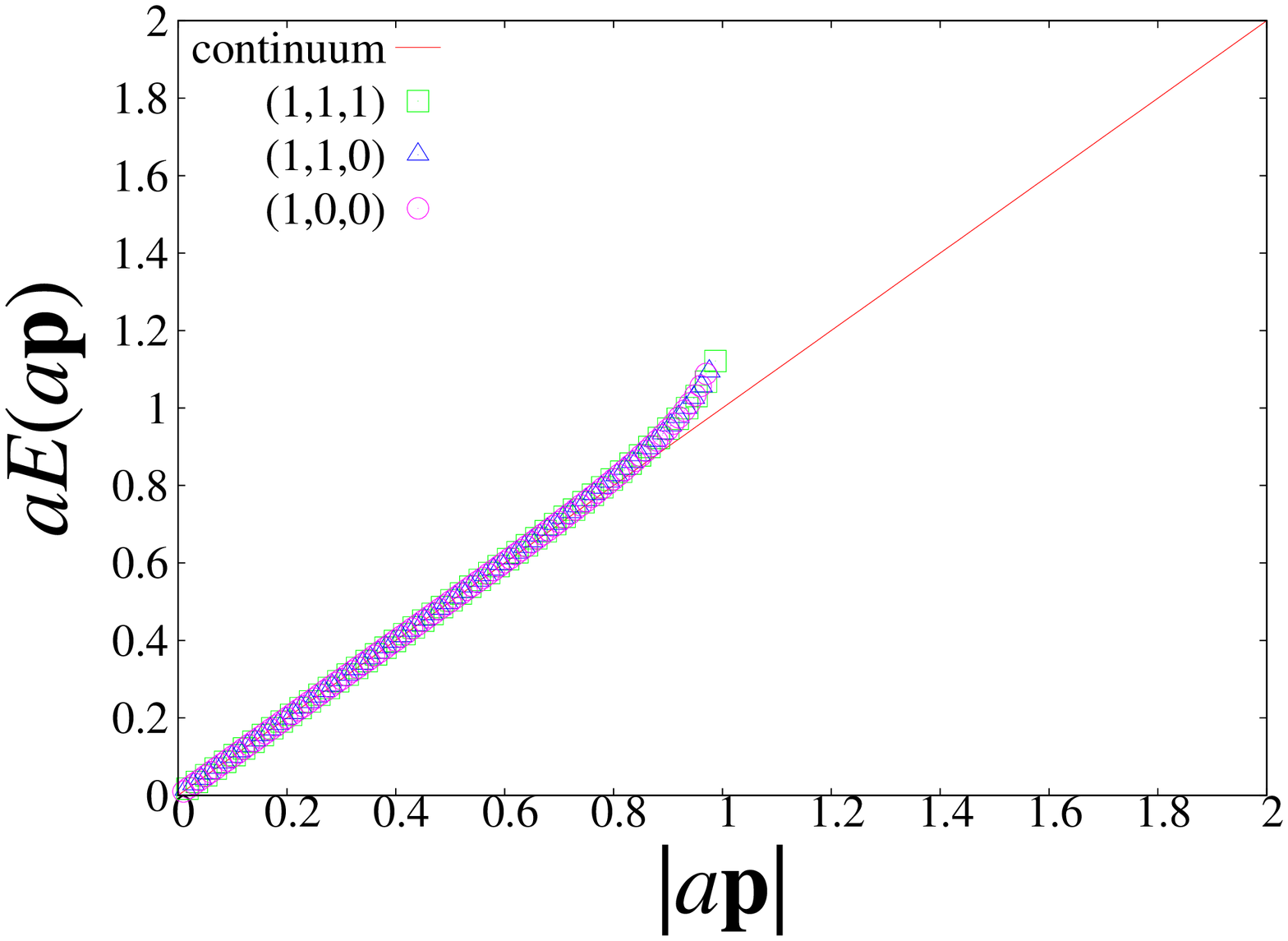}
  \includegraphics[scale=0.33]{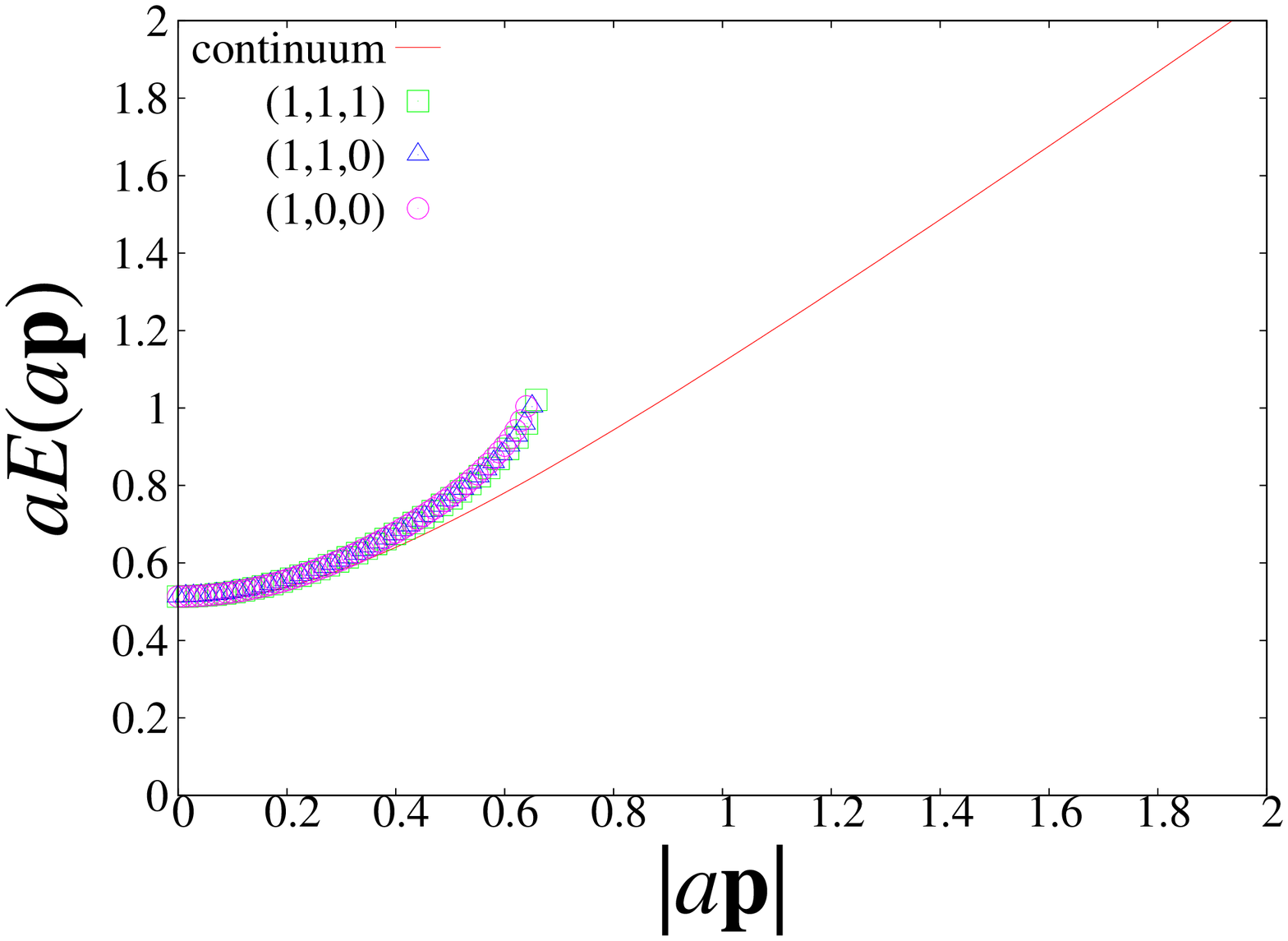}
  \caption{Same as Figure~\ref{fig:dispersion_Wilson}, but for the
    D34 action.}
  \label{fig:dis-d34}
\end{figure}

The dispersion relation for the D34 action is shown in Figure~\ref{fig:dis-d34} for the massless (left panel) and massive (right) cases. (We take $c_{D34}=1/6$ as in \cite{Alford:1996nx}.) Although they closely follow the continuum curve for small $a\bm{p}$, the solution disappears beyond $|a\bm{p}|\sim 1$. It is understood that the solution of the equation (\ref{eq:pole}) becomes complex, which is due to the lack of reflection positivity. It is potentially dangerous since the Wick rotation to the Minkowski
space is not doable in such a situation and one has to assume that the reflection positivity is recovered if the continuum limit is taken first.
It may have a practical problem that some instability occurs at relatively low momenta, especially for the massive case, as we discuss in the following sections.

\subsection{Improved Brillouin operator}
So far, we have shown that Brillouin fermion have some advantageous properties, even though it still contains the discretization effect of $O(a)$.
In the following, we attempt to eliminate these leading discretization errors by modifying the action.

Since the $O(a^2)$ error of $\nabla_\mu^{iso}$ keeps the rotational symmetry, its improvement is relatively simple. For instance, we may construct an improved Brillouin action as

\begin{equation}
  D^{imp} = \sum_\mu \gamma_\mu
  \left(1-\frac{a^2}{12}\triangle^{bri}\right)
  \nabla_\mu^{iso}
  \left(1-\frac{a^2}{12}\triangle^{bri}\right)
  + c_{imp} a^3(\triangle^{bri})^2,
  \label{eq:imp_bri_action}
\end{equation}
where we multiply the Laplacian operator from both sides of $\nabla_\mu^{iso}$ in order to preserve the $\gamma_5$-hermiticity property. The second term is simply squared with an arbitrary (positive) parameter $c_{imp}$.

\begin{figure}[tbp]
  \includegraphics[scale=0.33]{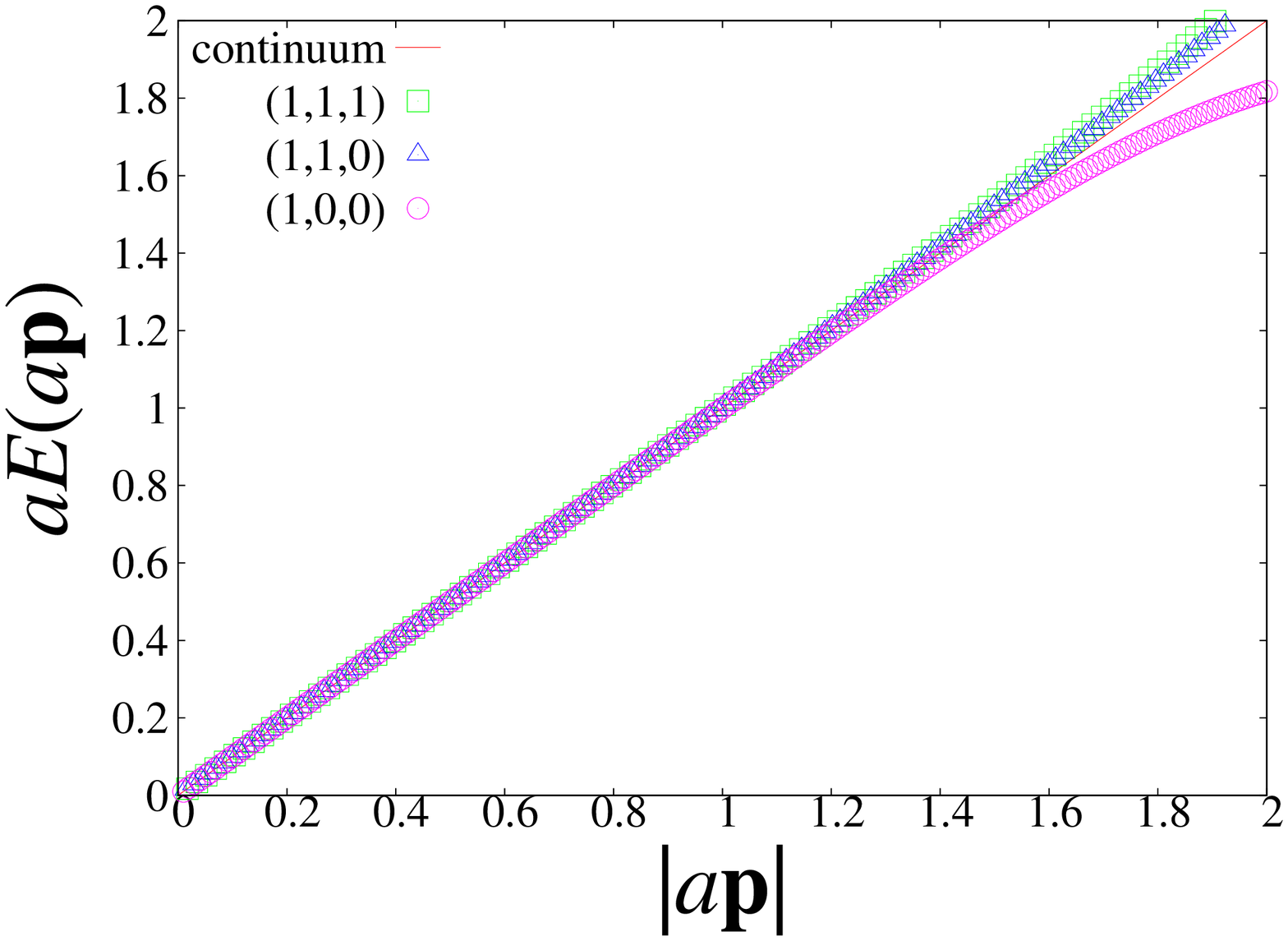}
  \includegraphics[scale=0.33]{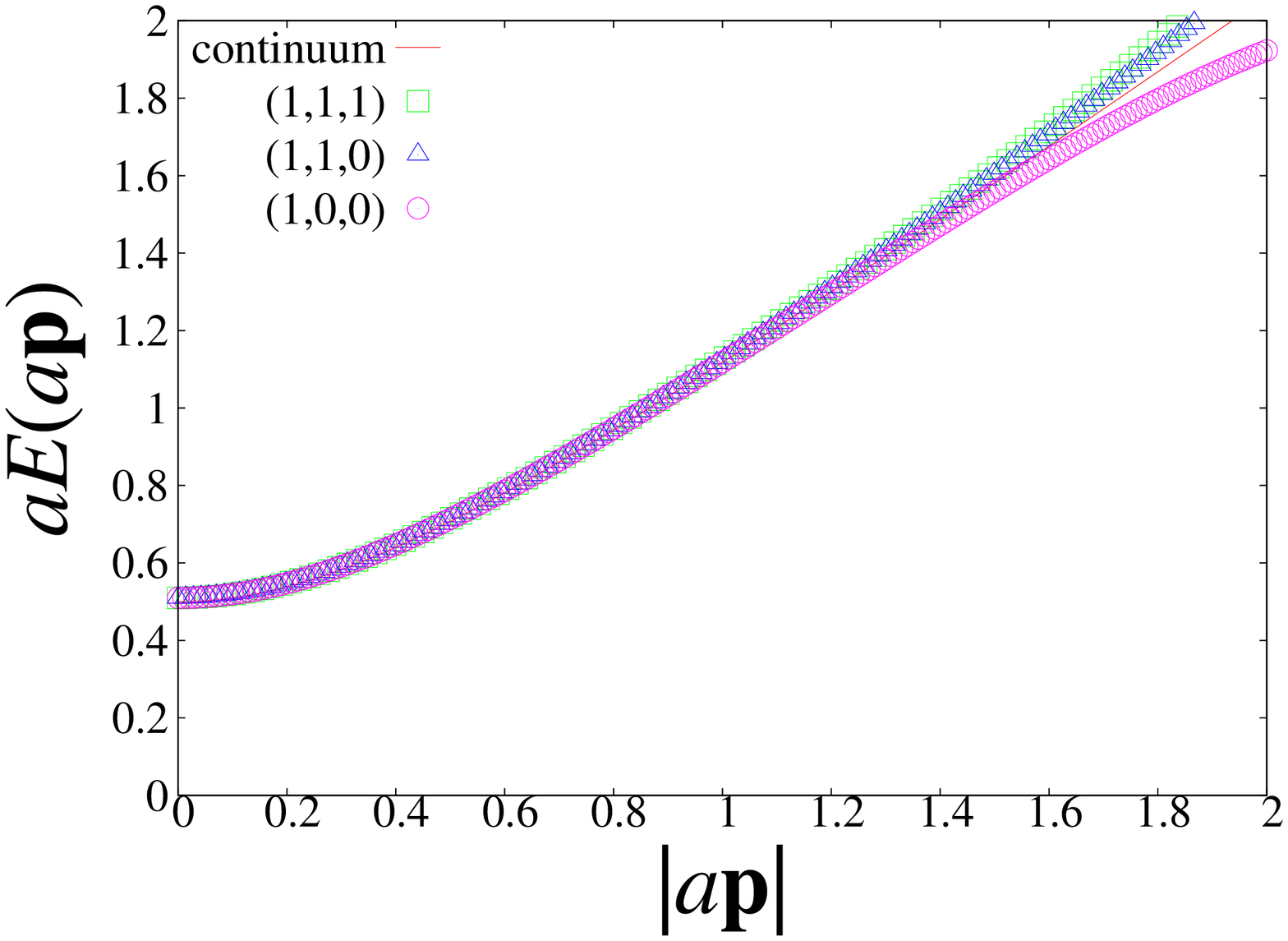}
  \caption{Same as Figure~\ref{fig:dispersion_Wilson}, but for the
    improved Brillouin action.}
  \label{fig:dis-ibr}
\end{figure}

This form of the improved action resembles the D34 action, but  using $\nabla_\mu^{iso}$ and $\triangle^{bri}$ as building blocks the energy-momentum dispersion relation is improved. As shown in Figure~\ref{fig:dis-ibr}, the dispersion relation gives a good approximation of the continuum up to $|a\bm{p}|\sim 1.5$. The Taylor expansion gives 

\begin{equation}
  (aE)^2(a\bm{p},am) = \left[(am)^2+2c_{imp}(am)^5 \right]
  + (a\bm{p})^2,
\end{equation}
which has the leading correction of $O(a^3)$ as expected and it does not contain the possible term of $O(a^3)$ that  violates the rotational symmetry.
This is because the building blocks  $\nabla_\mu^{iso}$ and $\triangle^{bri}$ themselves 
reduce the Lorentz violating effects.

\begin{figure}
  \centering
  \includegraphics[scale=0.4]{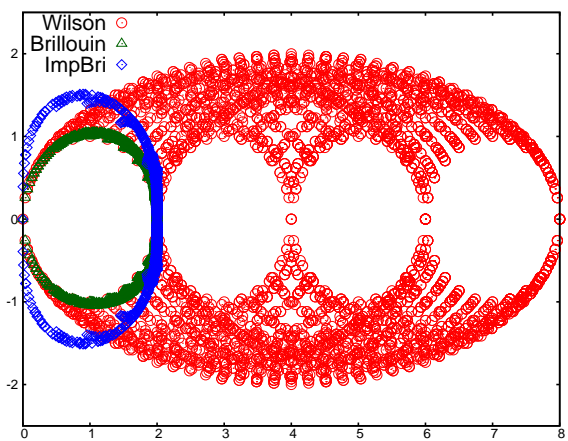}
  \caption{
    Eigenvalues of the lattice Dirac operators on the free background
    gauge field.
    The points show the eigenvalues of Wilson (red circles), Brillouin 
    (filled green triangles) and improved Brillouin (blue diamonds) fermion operators. 
  }
  \label{fig:Comparison-eigenvalues}
\end{figure}

The eigenvalue spectrum of the Dirac operator on the free background gauge field is shown in Figure~\ref{fig:Comparison-eigenvalues} for the improved Brillouin fermion (blue) together with those of Wilson (red) and Brillouin (green) fermions. The improved Brillouin eigenvalues form a circle structure similar to that of the Brillouin operator, but the circle is slightly squashed and pressed on the imaginary axis and approaches the continuum limit where the eigenvalues are purely imaginary.

\begin{figure}[tbp]
  \includegraphics[scale=0.33]{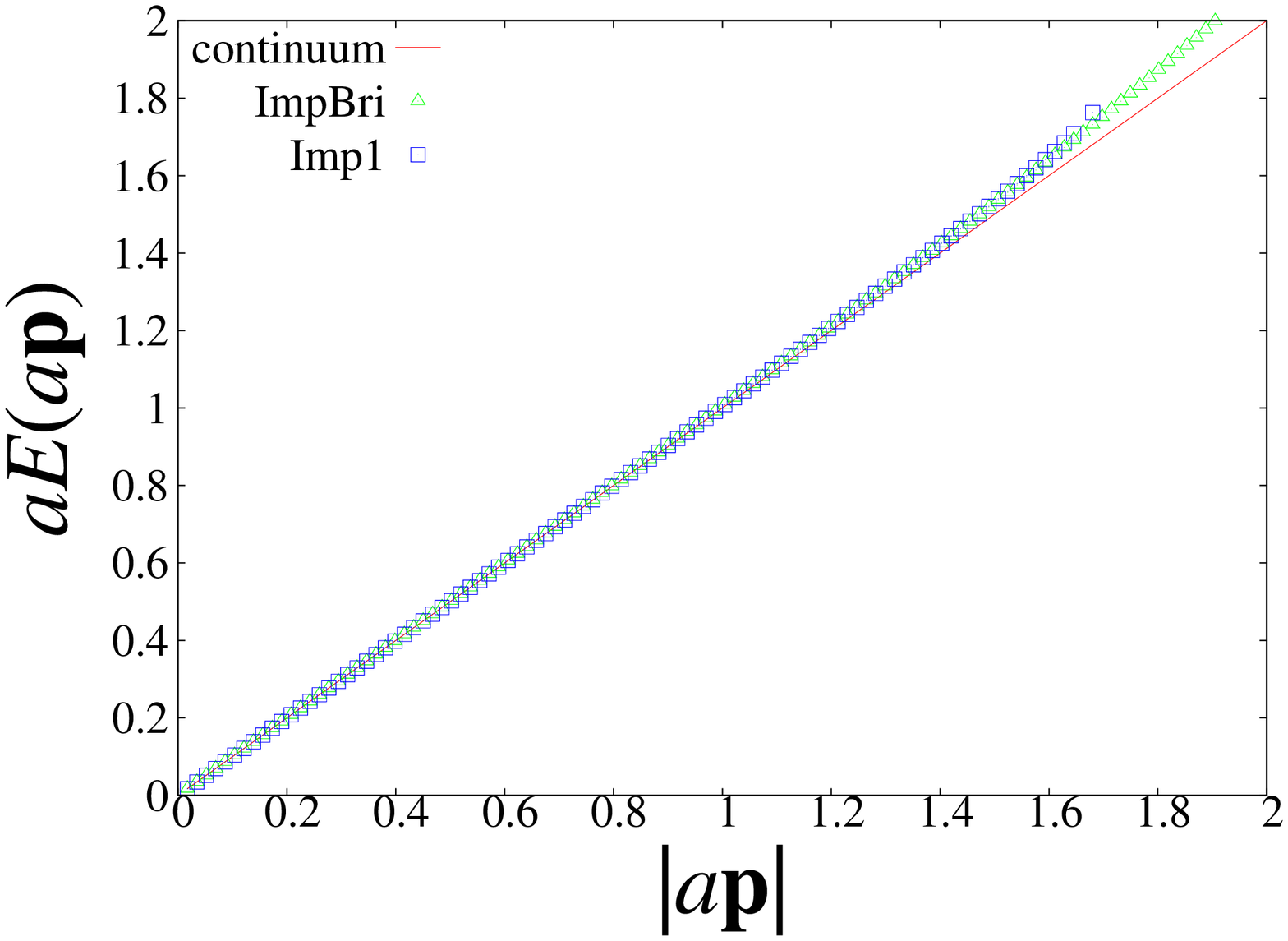}
  \includegraphics[scale=0.33]{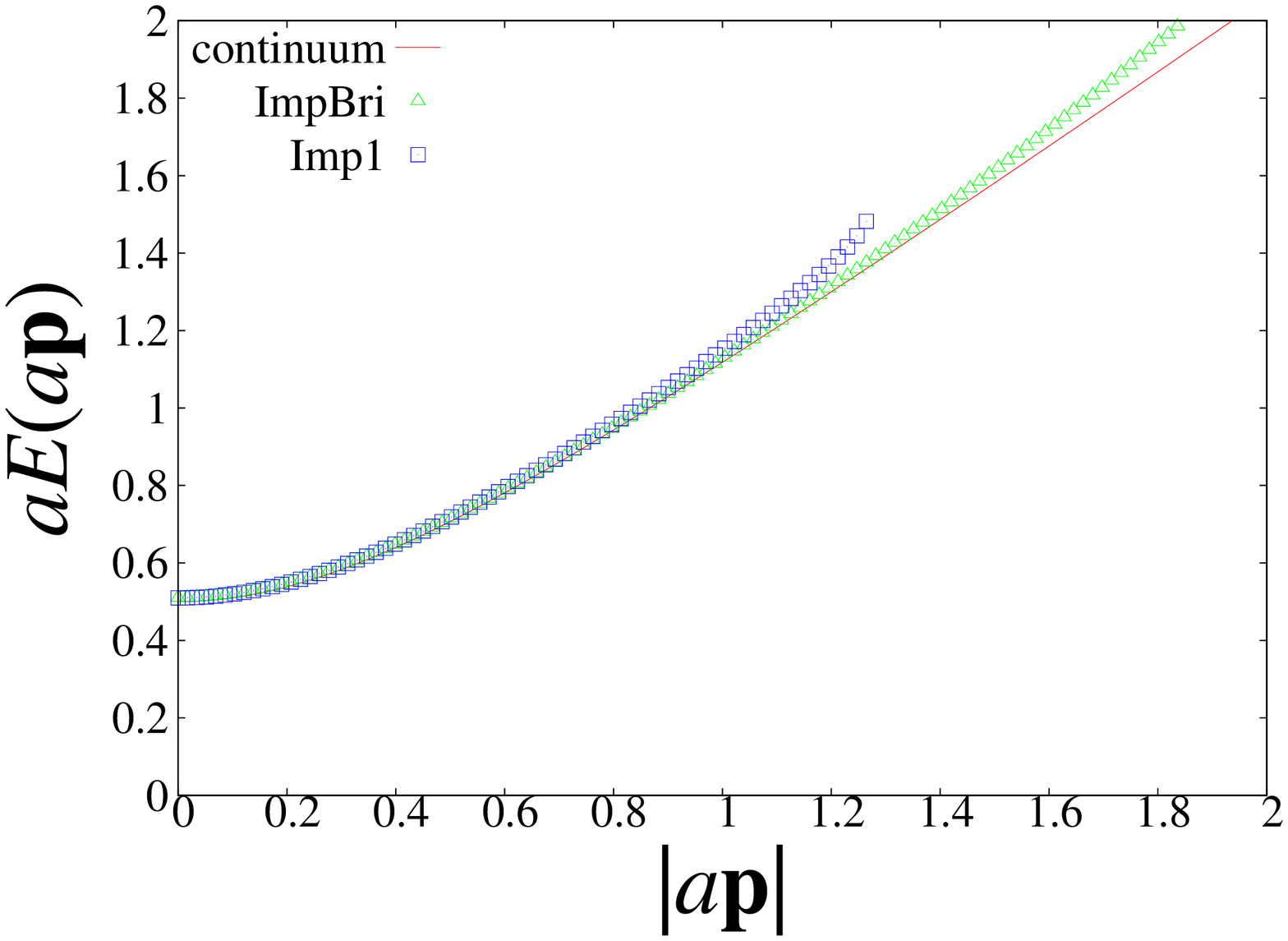}
  \caption{
    Same as Figure~\ref{fig:dispersion_Wilson}, but for the
    improved Brillouin action of reduced numerical cost
    (\ref{eq:imp_bri_action1}), which is shown by blue symbols.
    The improved Brillouin action of the original form
    (\ref{eq:imp_bri_action}) is also plotted in green for comparison.
  }
  \label{fig:dis_Dimp1}
\end{figure}

\begin{figure}[tbp]
  \centering
  \includegraphics[scale=0.4]{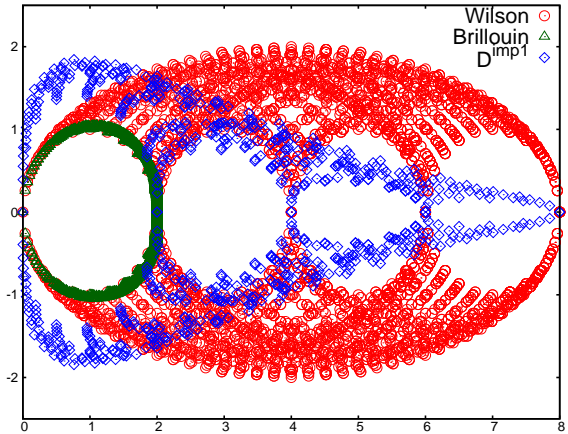}
  \caption{
    Same as Figure~\ref{fig:Comparison-eigenvalues}, but for $D^{imp1}$: the
    improved Brillouin action of reduced numerical cost
    (\ref{eq:imp_bri_action1}) with $c_{imp}=1/8$.
  }
  \label{fig:eigen_dimp1}
\end{figure}

The improved Brillouin operator $D^{imp}$ defined in (\ref{eq:imp_bri_action}) involves multiple applications of $\triangle^{bri}$, and therefore is numerically more expensive. Instead, we may consider a less expensive operator by using the standard operators for the terms introduced to cancel the $O(a^2)$ errors. Namely, we define

\begin{equation}
  D^{imp1} = \sum_\mu \gamma_{\mu}
  \left(1-\frac{1}{12}a^2\triangle^{std}\right) \nabla_\mu^{iso}
  \left(1-\frac{1}{12}a^2\triangle^{std}\right)
  + c_{imp} a^3 (\triangle^{std})^2,
  \label{eq:imp_bri_action1}
\end{equation}
where $\triangle^{std}$ is the standard lattice Laplacian operator. The energy-momentum dispersion relation for this modified operator is shown in Figure~\ref{fig:dis_Dimp1}. Unlike the original improved Brillouin action (\ref{eq:imp_bri_action}), the departure from the continuum relation is apparent already around $a|\bm{p}|\gtrsim$ 1.2. Furthermore, the eigenvalue distribution shown in Figure~\ref{fig:eigen_dimp1} demonstrates that the doubler spectrum splits as in the standard Wilson fermion. It is therefore expected that it requires more conjugate gradient iterations than the original improved Brillouin action to obtain the inverse. (See the discussions at the end of this section.)

\subsection{Overlap operators}
Since the Brillouin-Dirac operator has an eigenvalue distribution very similar to that of overlap fermions as demonstrated in Figure~\ref{fig:Eigenvalue-wil_vs_bri}, it may be an interesting option to use it as a kernel operator for the overlap-Dirac operator. Projection of eigenvalues to the unit circle in the complex plane would then require minimal numerical effort, {\it i.e.} the order of the Chebyshev polynomial or the Zolotarev rational function is relatively lower.

Another advantage of the overlap fermion is that the discretization effect of the massless Dirac operator is restricted to even powers of $a$ due to its exact chiral symmetry. For instance, if the standard Wilson-Dirac operator is used as a kernel of the overlap construction, the $O(a)$ error of Wilson fermions is eliminated and the leading error becomes $O(a^2)$. If the kernel operator is improved up to $O(a^2)$, then the discretization effect of the corresponding overlap operator starts from $O(a^4)$. The massless overlap-Dirac operator can be defined as

\begin{equation}
  D_{ov}(0)  = \frac{1}{Ra} \left[1+\frac{X}{\sqrt{X^{\dagger}X}}\right],
  \label{eq:Dov}
\end{equation}
where $X$ is a kernel operator with a large (negative) mass $\rho$ and $R$ is often taken to be proportional to the unit matrix.
Then, $D_{ov}$ satisfies the Ginsparg-Wilson relation $\{D_{ov},\gamma_5\}=RaD_{ov}\gamma_{5}D_{ov}$. Introducing a mass, the operator is modified to

\begin{equation}
  D_{ov}(m) = \left(1-\frac{am}{2\rho}\right) D_{ov}(0) + m.
\end{equation}

\begin{figure}[tbp]
  \includegraphics[scale=0.33]{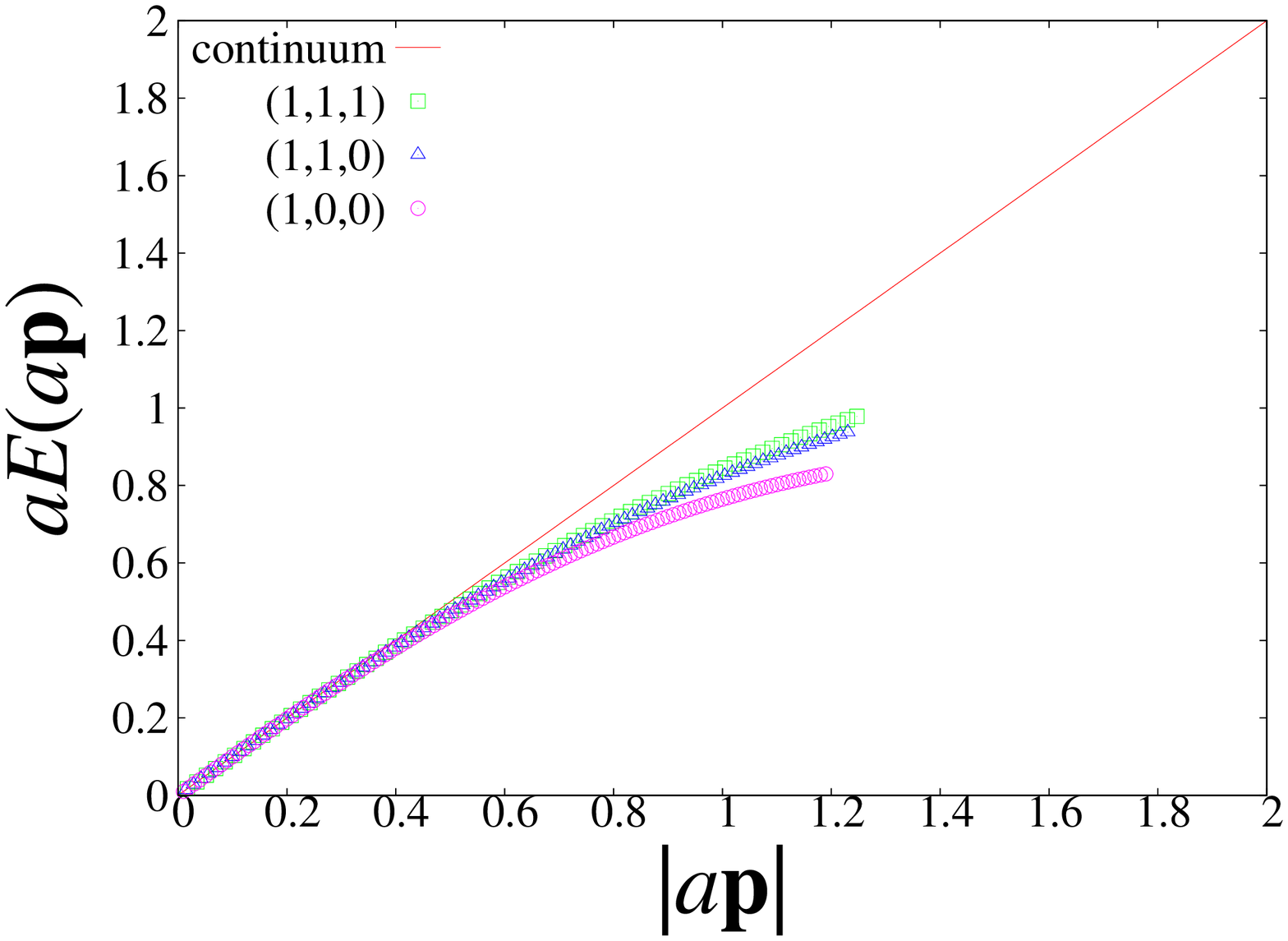}
  \includegraphics[scale=0.33]{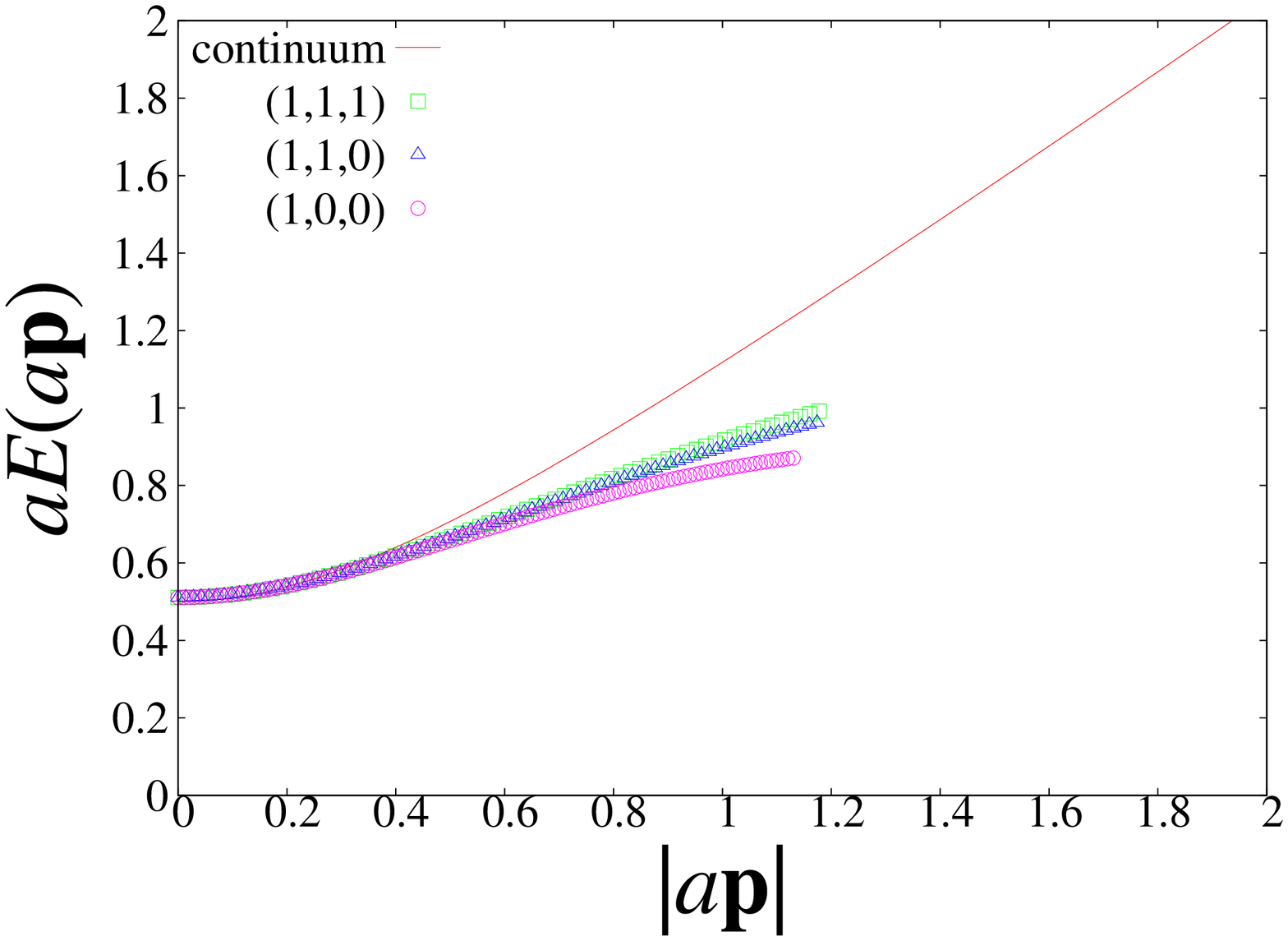}
  \caption{Same as Figure~\ref{fig:dispersion_Wilson}, but for the
    overlap fermion action with the standard Wilson kernel at
    $\rho=1$.
  }
  \label{fig:dis-ov-Wil}
\end{figure}

\begin{figure}[tbp]
  \includegraphics[scale=0.33]{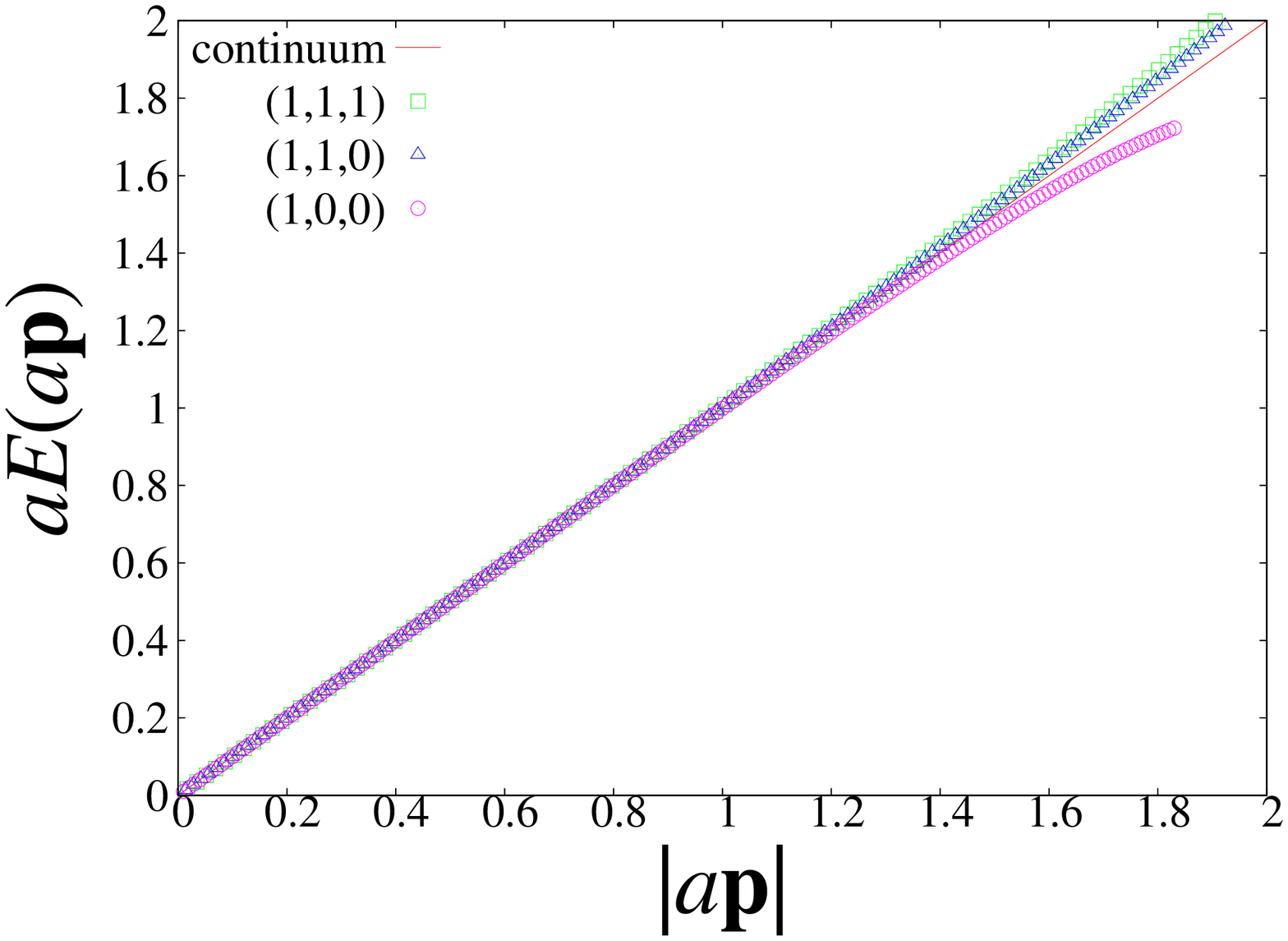}
  \includegraphics[scale=0.33]{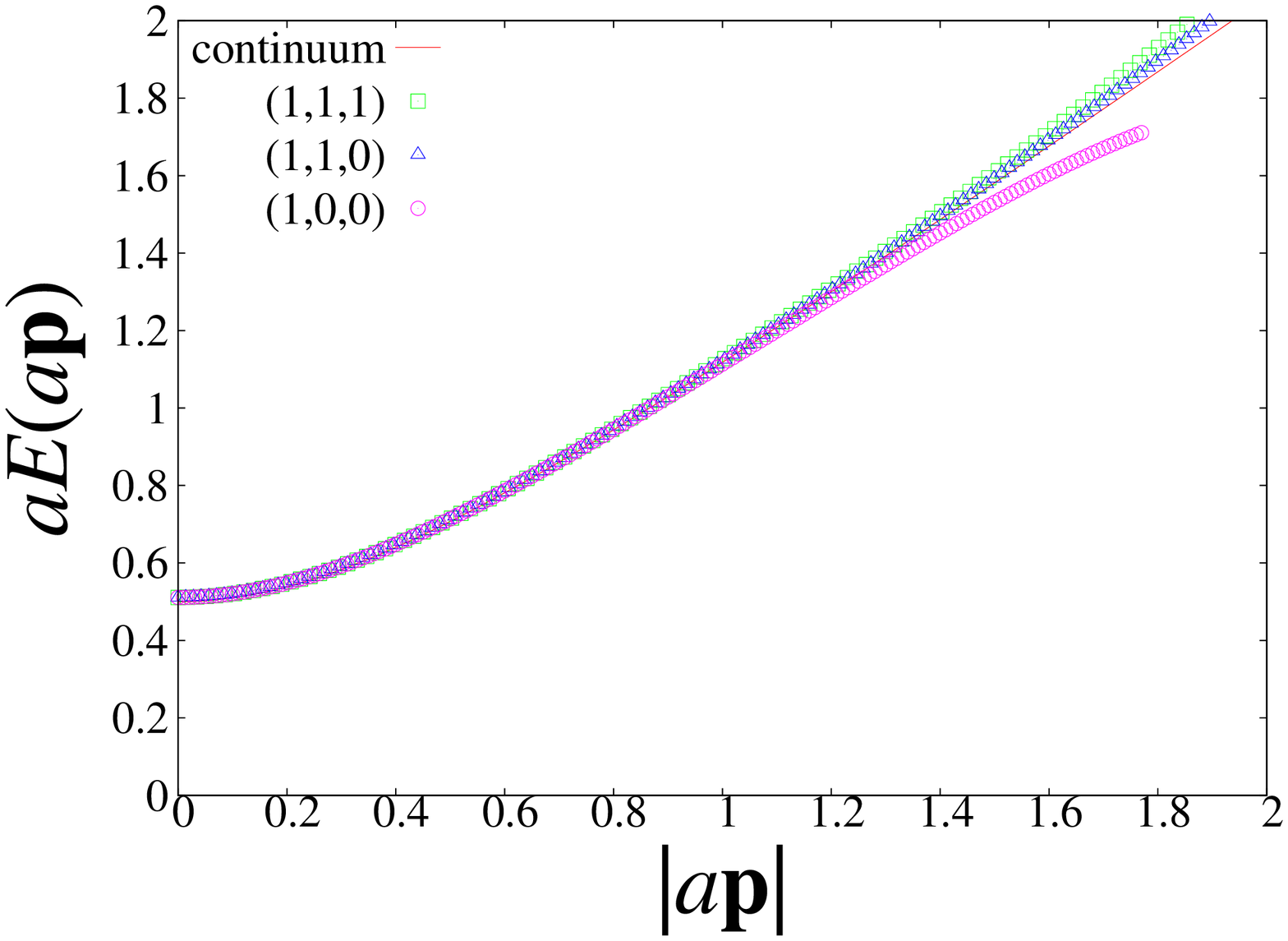}
  \caption{Same as Figure~\ref{fig:dispersion_Wilson}, but for the
    overlap fermion action with the Brillouin kernel at
    $\rho=1$.
  }
  \label{fig:dis-ov-Bri}
\end{figure}

It is straight-forward to write down the propagator and solve the pole to obtain the energy-momentum dispersion relation. With the standard Wilson kernel and $\rho=1$, the relation at $a\bm{p}=\bm{0}$ is

\begin{equation}
  (aE)^2 = (am)^2 + \frac{1}{6}\left( am \right)^4.
\end{equation}
This implies that the leading discretization effect is indeed $O(a^2)$. For finite momenta, we plot the dispersion relation in Figure~\ref{fig:dis-ov-Wil}. One can see that the dispersion relation is very similar to that of the kernel operator, which is in this case the Wilson-Dirac operator, shown in Figure~\ref{fig:dispersion_Wilson}. With the Brillouin operator as a kernel, the dispersion relation is improved as shown in Figure~\ref{fig:dis-ov-Bri}.

\begin{figure}[tbp]
  \includegraphics[scale=0.33]{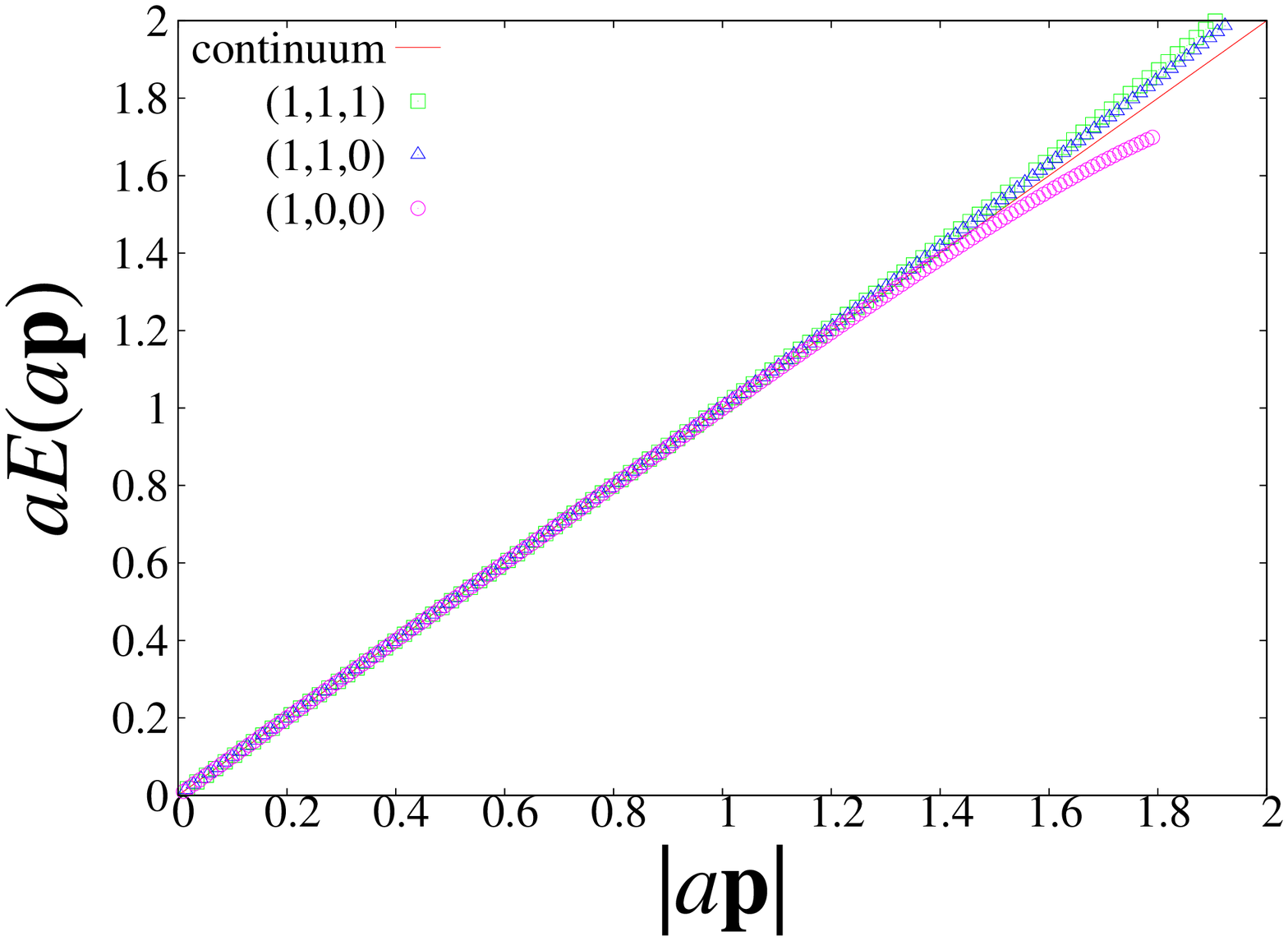}
  \includegraphics[scale=0.33]{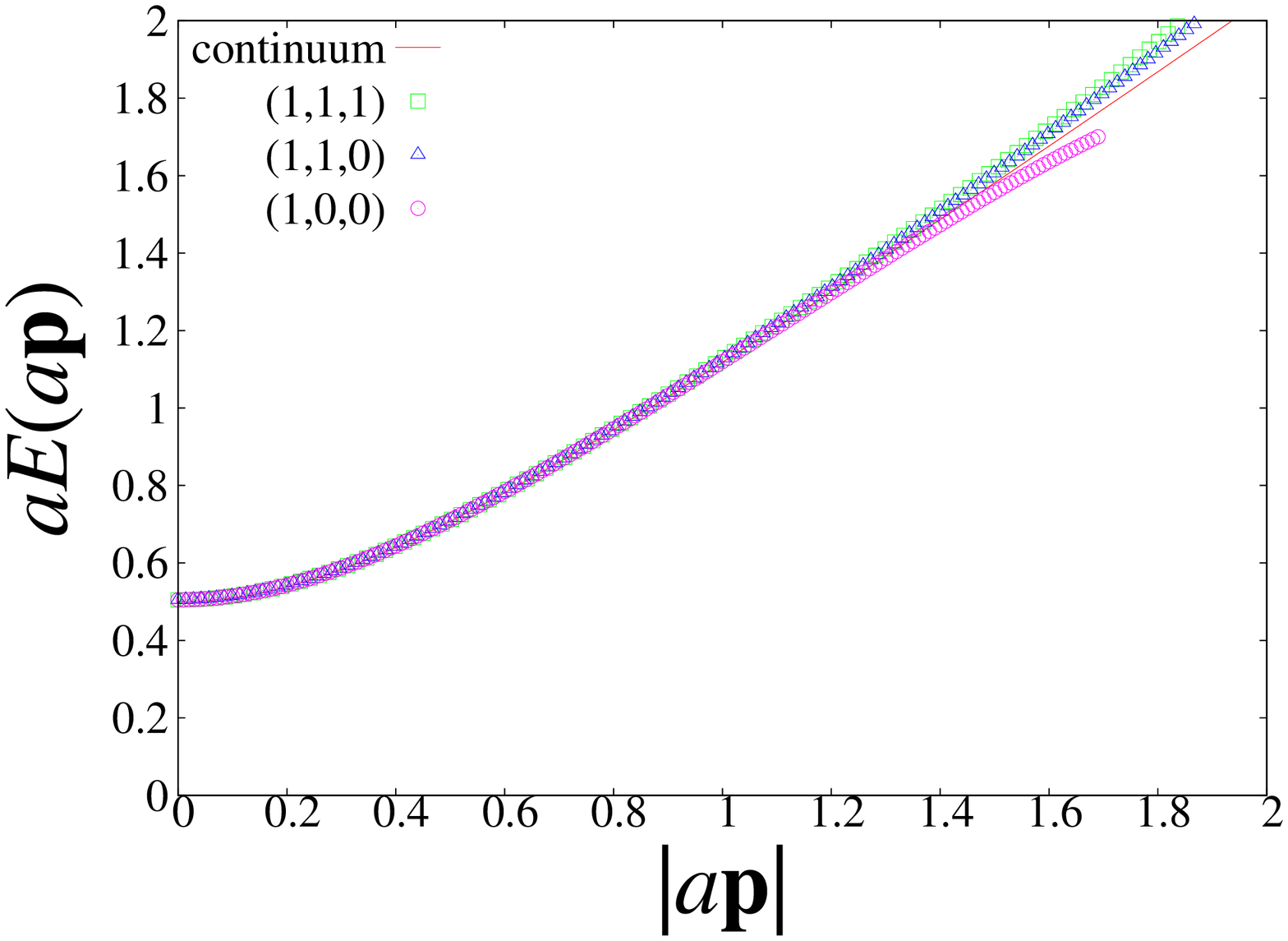}
  \caption{Same as Figure~\ref{fig:dispersion_Wilson}, but for the
    improved overlap fermion action with the improved Brillouin kernel
    at $\rho=1$.
  }
  \label{fig:dis-impov-impbri}
\end{figure}

Improving the overlap fermion action beyond the $O(a^2)$ discretization effects, one has to modify the construction of the overlap operator of (\ref{eq:Dov}), because the Ginsparg-Wilson relation of the form $\{D_{ov},\gamma_5\}=RaD_{ov}\gamma_{5}D_{ov}$ (with a constant $R$) already includes $O(a^2)$ effects. A possible modification is \cite{Ikeda:2009mv}

\begin{equation}
  D_{ov}^{imp}(m) 
  = m + \left(1-\frac{am}{2\rho}\right) D_{ov} + 
  \frac{a^2}{2\rho} D_{ov}^{\dagger}D_{ov}^2,
\end{equation}
where $D_{ov}$ is that of (\ref{eq:Dov}). In order to eliminate the $O(a^2)$ effects, it has to be used with an $O(a^2)$ improved kernel operator.  We calculate the dispersion relation for this improved overlap-Dirac operator with the improved Brillouin operator as a kernel. The result is shown in Figure~\ref{fig:dis-impov-impbri},
where we observe that a good approximation for the dispersion relation is maintained up to $|a\bm{p}|\sim \pi/2$. At zero spatial momentum, the relation $E=m$ is satisfied up to an error of $O(a^4)$.

The overlap operator (\ref{eq:Dov}) is usually constructed using a rational approximation, which is numerically expensive. The number of terms to be included in the rational approximation depends on the range of eigenvalues to be treated and on the desired precision.  When the kernel operator is already close to the overlap operator as in the case with the Brillouin operator, it is expected that a minimal order of the rational function would achieve a sufficient level of approximation. This property is still to be confirmed with actual numerical calculations.

\subsection{Numerical cost}
Although the advantage of the Brillouin-type Dirac operators is clear, its numerical cost is substantially higher than that of the standard (or improved) Wilson fermion action. This is simply because the isotropic derivative and the Brillouin Laplacian involves an interaction to $3^4-1=80$ neighboring points, which is ten times larger than the number of nearest neighbor points, $4\times 2=8$. Moreover, the reduction of numerical operation by a factor of two through taking advantage of the special $\gamma$ matrix combination $1\pm\gamma_\mu$ works only for the Wilson fermion. Therefore, we expect at least twenty times larger computational costs for the Brillouin operator, and in practice it is several times more, especially when we use the $O(a^2)$-improved version in (\ref{eq:imp_bri_action}). Therefore, in practical applications the improved Brillouin fermion could be used only for heavy quarks, for which the fermion matrix inversion can be carried out with small number of conjugate gradient iterations.

\begin{figure}[tb]
  \centering
  \includegraphics[scale=0.33]{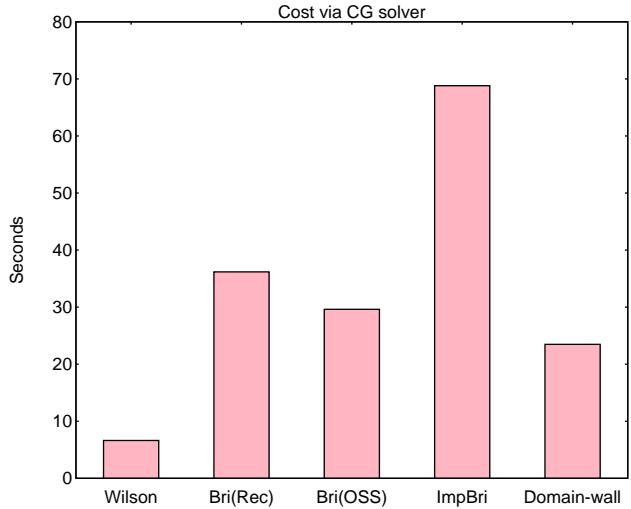}
  \caption{
    Numerical cost for various Dirac operators.
    Elapsed time (in sec) to solve the heavy quark propagator using the conjugate gradient method is plotted.
    The pseudo-scalar meson mass $m_{PS}$ is roughly tuned to 3.0~GeV.
    Results are plotted for Wilson fermions, Brillouin fermions (Rec and OSS implementations), the improved Brillouin fermion, and
    the domain-wall fermion.
    For the domain-wall fermion, the lattice size in the fifth direction is taken as $L_s=8$.
  }
  \label{fig:comp_numerical_cost}
\end{figure}

In Figure~\ref{fig:comp_numerical_cost} we compare the numerical costs for  various Dirac operators by measuring the elapsed time to solve the heavy quark propagator corresponding to the pseudo-scalar meson mass $m_{PS}\simeq$ 3.0~GeV. A quenched $16^3\times 32$ lattice of $1/a\simeq$ 2.0~GeV is chosen for the
test and the numerical computation is done on a 32-node partition of the IBM Blue Gene/Q machine. For the solver we employ the conjugate gradient method for $D^\dagger(m)D(m)$. From Figure~\ref{fig:comp_numerical_cost} we can see that the Brillouin fermion takes only five-times more time than Wilson fermions does, despite the above expectation. Likewise, the improved Brillouin fermion is only ten-times slower than Wilson fermions.
For the Brillouin fermion, two implementations are attempted, {\it i.e.} the overall smearing strategy (OSS) and the recursive formula (Rec) as described in the Appendix~\ref{sec:hops}. The OSS implementation has an additional cost, which we did not account for here, due to an uncounted cost to setup diagonal gauge links.

We also notice that the performance of the computation on the Blue Gene/Q is different for different fermion actions. In our implementation, the number of floating point operation per second (GFlops) per node is about 4.0 for Wilson fermions, while that for other actions is around 10, which is compared to the peak performance 200~GFlops, because they are more compute-intensive. The elapsed time is thus relatively shorter for the actions other than Wilson. (In other applications, the JLQCD collaboration uses a highly optimized code for the Wilson fermion, which performs much better than 15~GFlops depending on the condition, but for the comparison in Figure~\ref{fig:comp_numerical_cost} we used a more primitive version of the Wilson-Dirac operator in order to make a fair comparison. Optimization of the Brillouin operators is yet to be done.)

\begin{figure}[tb]
  \centering
  \includegraphics[scale=0.33]{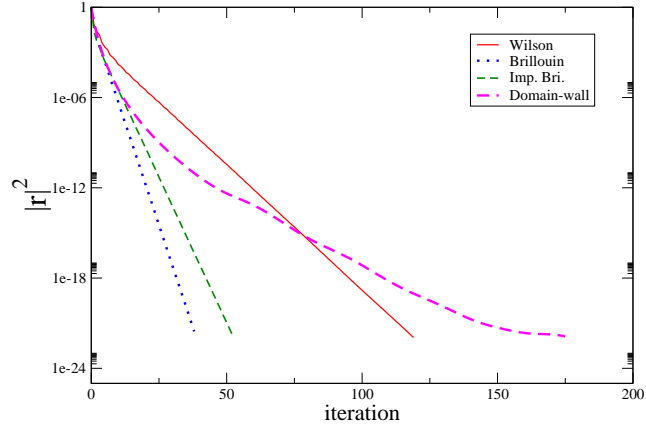}
  \caption{
    Squared norm of the residual vector at every conjugate gradient
    iteration steps. 
    Data for Wilson fermions (solid), Brillouin fermions (dotted), improved Brillouin fermion (thin dashed) and the domain-wall fermion (thick dashed) are shown.
  }
  \label{fig:history_CG}
\end{figure}

This relative speed-up of the Brillouin fermion is explained by  the number of conjugate gradient iterations to converge. Figure~\ref{fig:history_CG} shows the squared norm of the residual vector at every conjugate gradient iteration steps for Wilson fermions, Brillouin fermions, the improved Brillouin fermion and the domain-wall fermion. The number of iterations is clearly smaller for the Brillouin-type fermions by more than a factor of two. This explains why the Brillouin-type fermions are not as slow as we naively expect. It comes from the fact that the largest eigenvalue of $|D(0)|$ is 2 for the Brillouin operator rather than 8 of Wilson fermions. The condition number of the matrix $D(m)$ is thus four-times smaller for the Brillouin-type fermion.

\section{Limitation on the quark mass for the improved actions}
\label{sec:heavy_problem}

\begin{figure}[tb]
  \centering
  \includegraphics[scale=0.50]{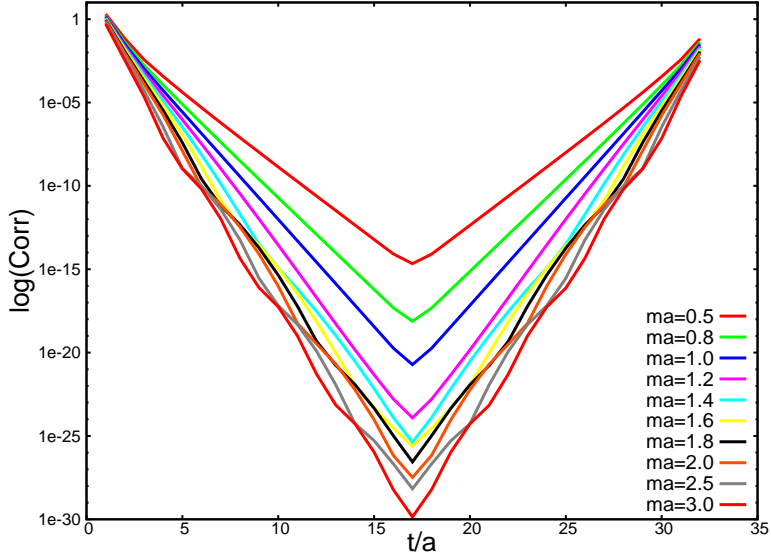}
  \caption{
    Heavy pseudo-scalar meson correlators for various heavy quark masses. 
    The statistical error is smaller than the thickness of lines.
    Lines show the correlators of different masses: $am$ = 0.5, 0.8, 1.0, 1.2, 1.4, 1.6, 1.8, 2.0, 2.5 and 3.0 from top to bottom.
  }
  \label{fig:Oscillating-euclidean-propagator}
\end{figure}

In general, higher derivative terms may give rise to some unphysical poles \cite{Alford:1996nx}, which are sometimes called ``ghost'' or ``lattice ghost.'' If the mass of the ghosts are sufficiently large, no physical effect can be observed, but once they come close to the physical pole, ghosts may distort the physical solution. Practically, it appears as an oscillatory behavior of the Euclidean correlator. For instance, Figure~\ref{fig:Oscillating-euclidean-propagator} shows the pseudo-scalar meson correlators calculated with improved Brillouin fermions at various quark masses up to $am=3.0$. For $am\gtrsim$ 1.5, one finds that the correlator is no longer a simple exponential function but is oscillating. Once this happens, the Wick rotation back to the Minkowski space can not be performed. This problem typically shows up only when the improvement including next-to-nearest neighbor interactions is introduced and when the bare quark mass $am$ is large. Therefore there is an upper limit on $am$ to avoid such sickness. One has to be careful, because the problem may be hidden even when the resulting correlation function does not show the oscillatory behavior.

\begin{figure}[tb]
  \centering
  \includegraphics[scale=0.50]{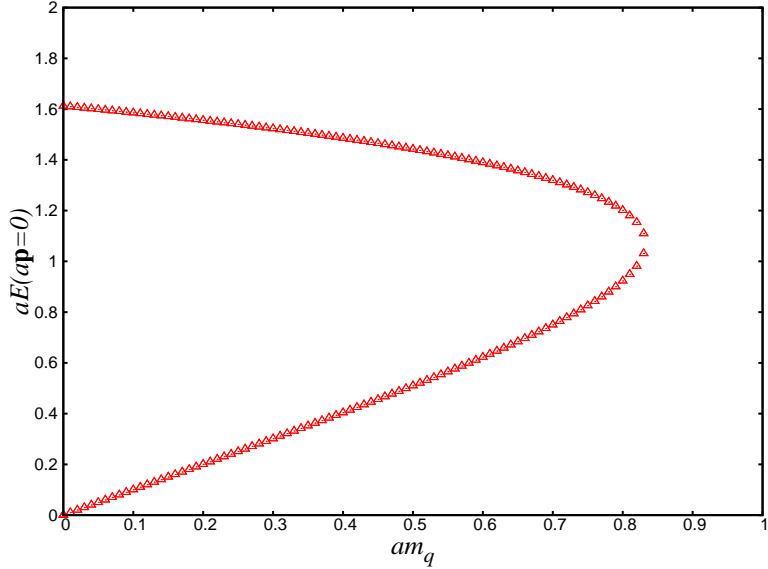}
  \caption{
    Energy of the physical pole (lower curve) and of the ghost (upper curve) as a function of bare quark mass $am$.
  }
  \label{fig:real_energy}
\end{figure}

At tree-level, we calculate the energy of the physical pole as well as that corresponding to the lightest doubler. Figure~\ref{fig:real_energy} shows those for zero spatial momentum as a function of $am$. The energy from the physical pole follows the expectation $E(|\bm{p}|=0)\simeq m$, up to $am$ = 0.6--0.7. On the other hand, the doubler mass slightly decreases for larger $am$ and comes close to the physical pole near $am\simeq$ 0.84. Beyond this value, the two poles merge and transform to a complex-conjugate pair, which indicates the ``ghost'' as discussed above. (The position of the ``merging'' point depends on the details of the action, and can in fact be slightly pushed to $am\simeq$ 0.97 for $c_{imp}=1/16$ instead of $c_{imp}=1/8$.)

In order to avoid unwanted effects due to the doubles and ghosts, we need to keep their energy sufficiently higher than the physical mode. By requiring a ``gap'' of $O(1/a)$, the upper limit of the heavy quark mass would be 0.5--0.6, according to the tree-level analysis shown in Figure~\ref{fig:real_energy}. The effect of the doublers/ghosts on non-perturbative physical observables may appear as a larger scaling violation. Such a symptom will be discussed in the end of the next section.

\section{Scaling studies on quenched configurations}
\label{sec:scaling_study}
In this section, we describe a non-perturbative scaling study of the improved Brillouin fermion as well as the standard Wilson and domain-wall fermions towards the continuum limit. We monitor the energy-momentum dispersion relation and hyperfine splitting of charmonium-like heavy-heavy mesons on a set of quenched lattices of inverse lattice spacing between 2.0 and 5.6~GeV.

\subsection{Lattice parameters}

We generate a set of SU(3) quenched lattices with the tree-level Symanzik gauge action at $\beta$ = 4.41, 4.66, 4.89 and 5.20 as summarized in Table~\ref{tab:parameters}. 
All lattices have a roughly constant spatial volume $L^3$ with $L\simeq$ 1.6~fm. These lattices have inverse lattice spacing between 2.0~GeV and 5.6~GeV. The lattice spacing is fixed through the gradient flow using an input $w_0$ = 0.176(2)~fm \cite{Borsanyi:2012zs}. The lattice spacing determined from the Sommer scale $r_0$ = 0.49~fm is also listed for three coarser lattices. All ensembles are generated with the heatbath algorithm and the measurement is carried out on gauge configurations separated by $N_{sep}$ heatbath sweeps, so that the auto-correlation can be safely neglected.  The number of statistical samples is around 100 for each $\beta$ value except for the finest lattice where we have 36 independent gauge configurations. 
The link smearing procedure is applied on the gauge configurations before using for the measurement of the heavy-heavy meson correlators (except for that of the ``unsmeared'' domain-wall fermion, as described below). To be explicit, we employ stout smearing \cite{Morningstar:2003gk} with a parameter $\alpha=0.1$ and repeat it three times.

\begin{table}[tbp]
  \centering
  \begin{tabular}{cccccc}
    \hline\hline 
    $L/a$ & $\beta$ & $N_{sep}$ & $a^{-1}$ [GeV] & $a^{-1}$ [GeV] & $L$ [fm]\\
    \hline 
    16 & 4.41 & 100 & 2.00(07) & 2.06(04) &1.579(55) \\
    \hline 
    24 & 4.66 & 200 & 2.81(09) & 2.89(15) &1.686(52) \\
    \hline 
    32 & 4.89 & 500 & 3.80(12) & 3.81(09) &1.664(51) \\
    \hline 
    48 & 5.20 & 40,000 & 5.64(22) & N/A &1.683(64) \\
    \hline\hline 
  \end{tabular}
  \caption{
    Quenched lattices used in this study.
    Temporal lattice size is always $T/a = 2L/a$.
    The fourth column shows $a^{-1}$ determined from the gradient flow.  The fifth column is an estimate of the lattice scale from the Sommer scale $r_0 = $ 0.49 fm.
  }
  \label{tab:parameters}
\end{table}

We study the continuum scaling of the improved Brillouin fermion defined by (\ref{eq:imp_bri_action}) with $c_{imp}=1/8$. For comparison, we also employ the standard Wilson fermion and the M\"obius domain-wall fermion. For M\"obius domain-wall fermions, we chose two options: smear or unsmear the gauge links. M\"obius domain-wall fermions are essentially the same as the conventional domain-wall fermions, but they are designed to achieve much smaller violation of chiral symmetry at a fixed length $L_s$ in the fifth dimension \cite{Brower:2012vk}. We chose the scale factor $b_5+c_5$ to be 2.0 and $L_s$ = 8  with the domain-wall height $M_0$ = 1.0 for the smeared domain-wall fermion. The residual breaking of chiral symmetry in this setup is found to be $O(\mbox{1~MeV})$ \cite{Hashimoto:2014gta}.

In the following we first describe the measurements with the smeared gauge link. Another set of measurements with unsmeared domain-wall fermion is separately discussed below. We tune the quark masses so that pseudo-scalar meson masses become close to our target values $m_{PS}$ = 1.0, 1.5, 2.0, 2.5, 3.0 and 3.5~GeV for all three fermion formulations. The numerical results are interpolated to these target values before comparing the final results. Since the length of this interpolation is tiny, we only use a linear function between nearest two data points.

\begin{figure}[tbp]
  \centering
  \includegraphics[scale=0.3]{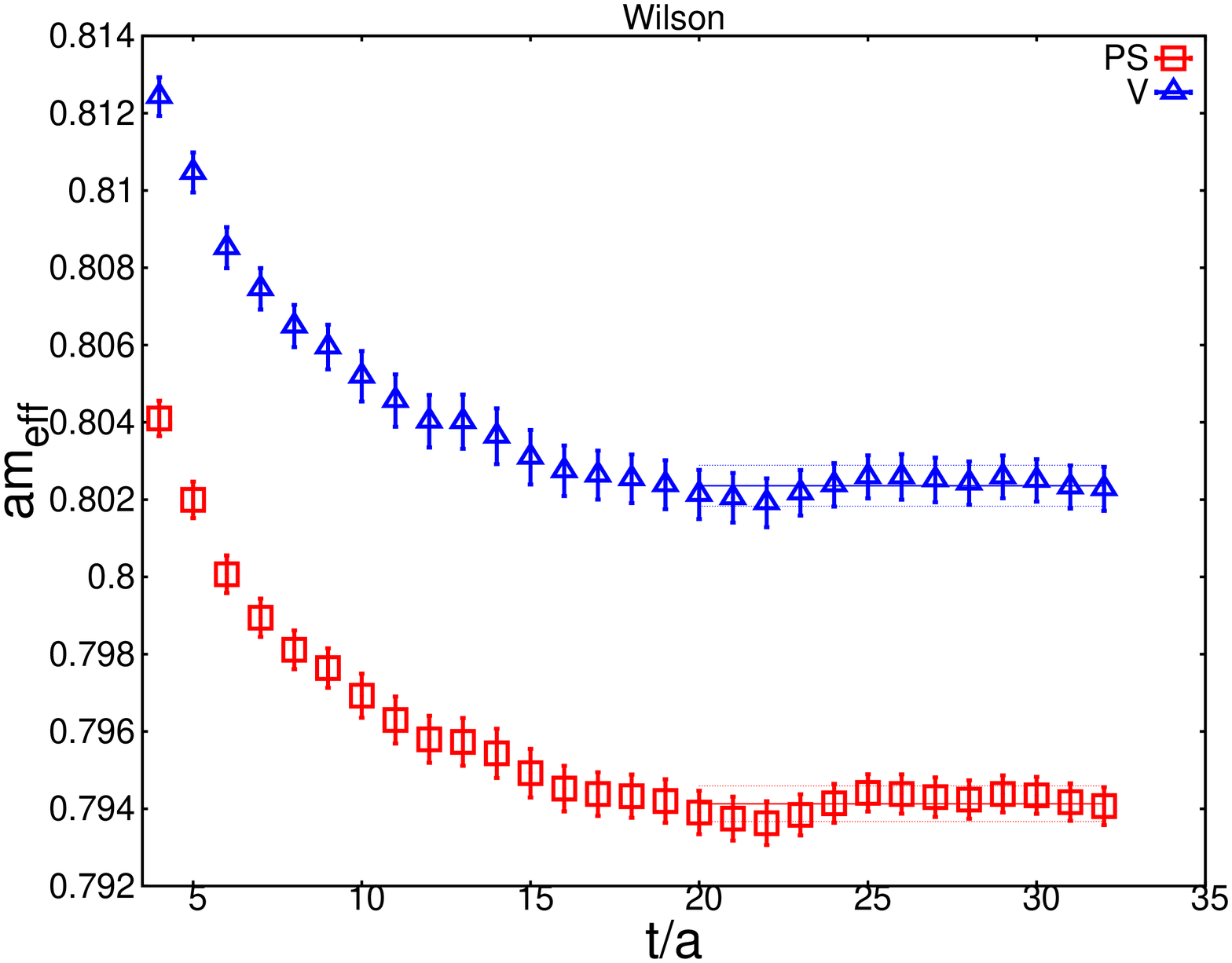}
  \includegraphics[scale=0.3]{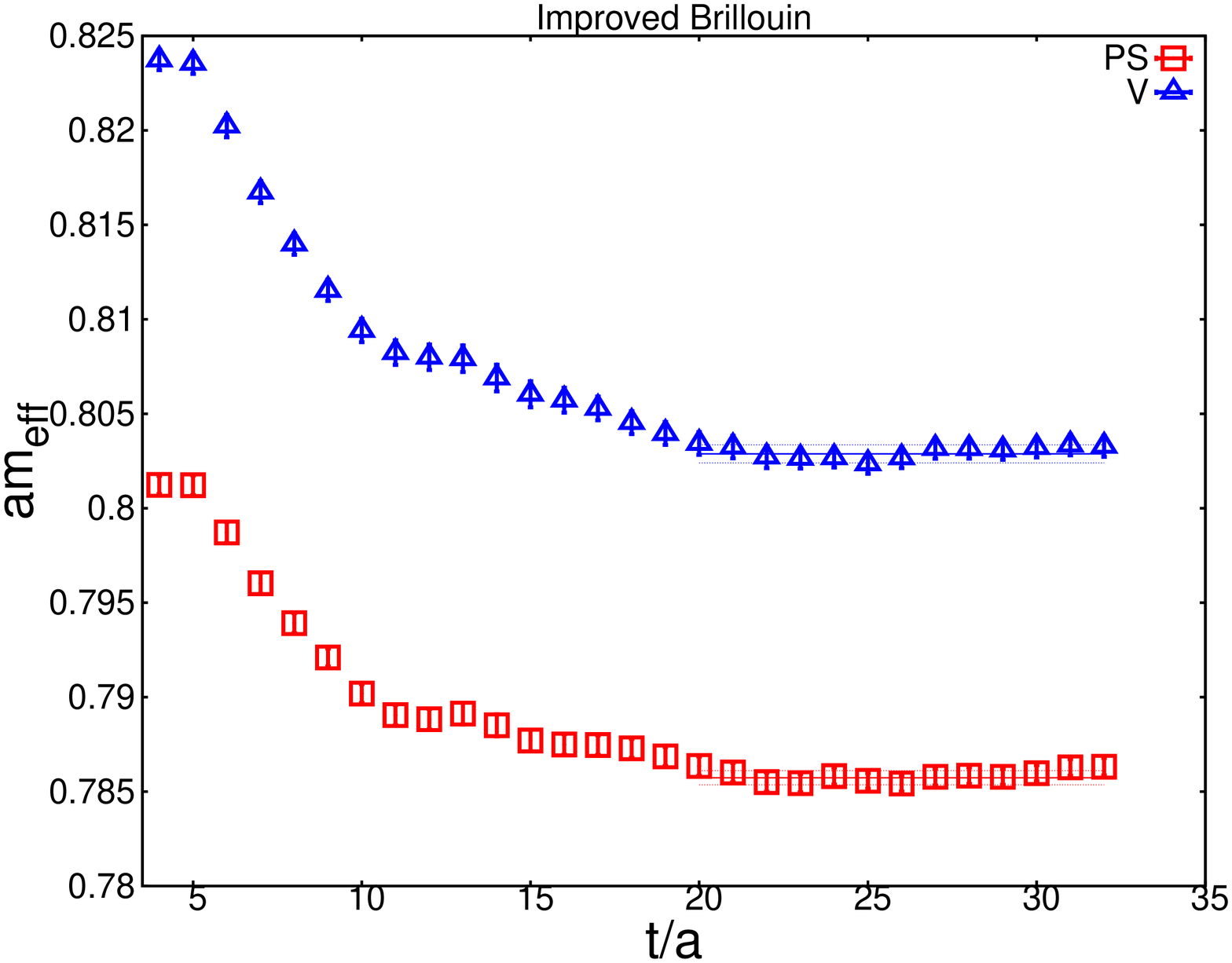}
  \includegraphics[scale=0.3]{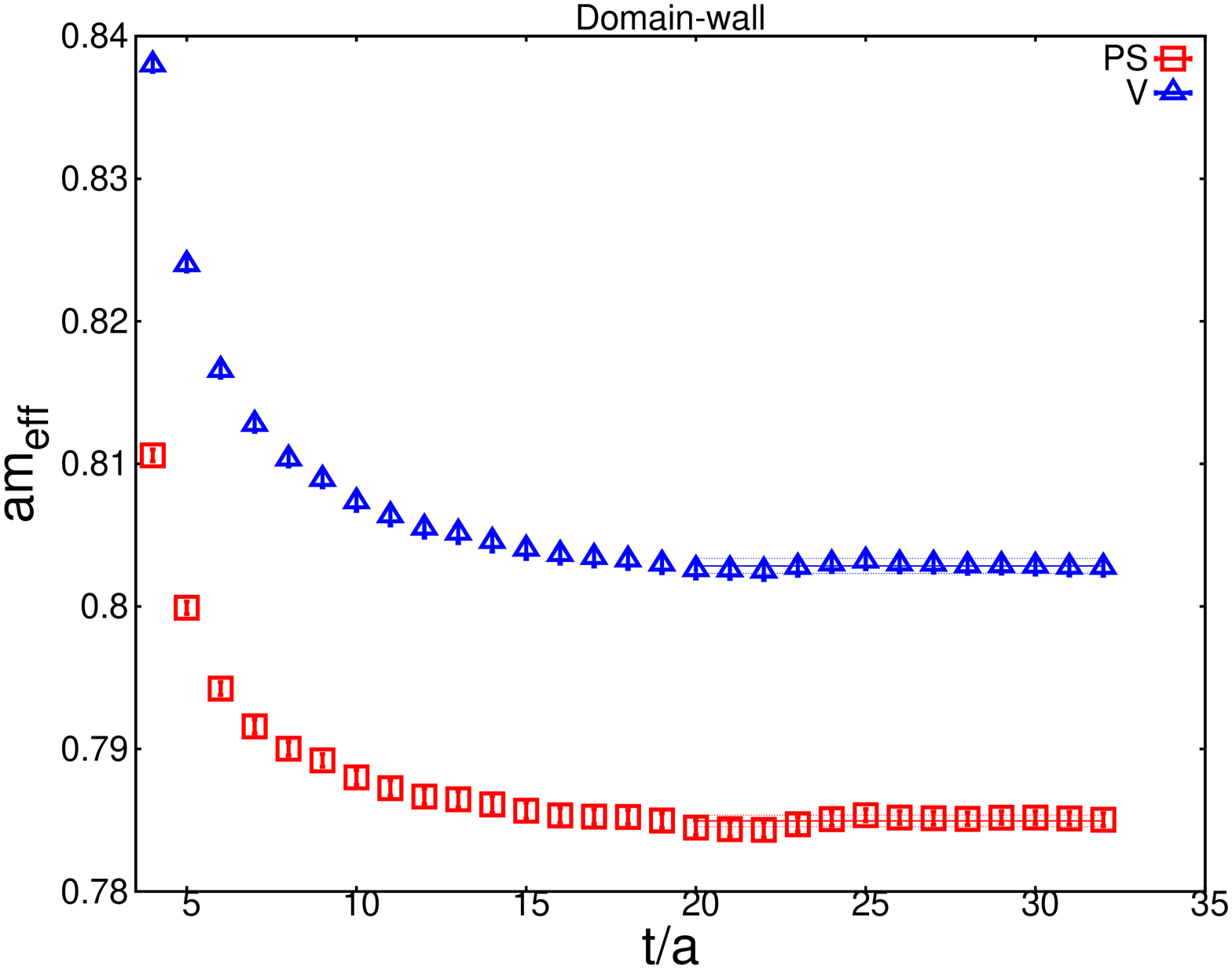}
  \caption{
    Effective mass of the pseudo-scalar (squares) and vector (triangles) 
    mesons for the mass parameter corresponding to $m_{PS}$ = 3.0~GeV. 
    Data for Wilson fermions (top), improved Brillouin fermions
    (middle) and smeared domain-wall fermions (bottom) at $\beta$
    = 4.89 are plotted. 
    Lines show the fit range and fitted value. 
  }
  \label{fig:effm_typ}
\end{figure}

We calculate heavy-heavy meson correlators from four different source points in the time direction. We smear the source with a function $e^{-\mu_{smr} r}$ with a parameter $\mu_{smr}$ tuned for each mass to obtain better saturation of the ground state. The gauge configurations are fixed to Coulomb gauge. The mass and smearing parameter are listed in Table~\ref{tab:param}. The effective masses for the ground state pseudoscalar and vector mesons are shown in Figure~\ref{fig:effm_typ}. The data corresponding to  $m_{PS}\simeq$ 3.0~GeV calculated on the $\beta$ = 4.89 lattice are taken as a typical example. We fit the lattice data with a single exponential function (plus the term representing the contribution from the other temporal direction) in a range $[t_{min},t_{max}]$ = [20,32]. To estimate the systematic error due to the fits, we repeat the fit with larger $t_{min}$'s until $t_{min}$ = 29 and take the variation of their central values as the size of systematic error. This similar procedure is applied for other $\beta$ values and for all masses. The fit results and associated error are also shown in the plots by horizontal lines.

In the case of the unsmeared M\"obius domain-wall fermion a slightly different strategy and set of parameters are chosen. 
Since the unsmeared gauge links are relatively rough, we take a larger value of $L_s$ ($L_s$ = 12) with $M_0$ = 1.6. 
We work with two different source types (point and complex $Z_2$-wall source \cite{Foster:1998vw}).
For each source type, we also calculate the quark propagator with a gauge-covariant Gaussian smearing applied on either source or sink (or both). 
We take many quark masses as described in Appendix~\ref{sec:param_unsmr}.
The results of the correlator fits are interpolated to the same reference pseudo-scalar masses in Table~\ref{tab:param}. This interpolation is carried out with two different ansatzes: a linear interpolation between the nearest two and a quadratic interpolation between the nearest three simulated data points. The spread of the central values between the two different approaches gives rise to a systematic error that has been taken into account in the analysis of the continuum extrapolation. 
In the data for the hyperfine splitting we see a variation of the central value by up to one sigma when varying the fit-range for the two point functions over a wide range. 
In order to remain on the conservative side we attach a systematic error of the size of the statistical error.

\begin{table}[tbp]
  \centering
  \begin{tabular}{cccccccc}
    \hline 
    $m_{PS}$ [GeV] & Dirac op. & 
    \multicolumn{2}{c}{$\beta=4.41$} & 
    \multicolumn{2}{c}{$\beta=4.66$} & 
    \multicolumn{2}{c}{$\beta=4.89$}\\
    \hline 
    & & $am$ & $\mu_{smr}$ & $am$ & $\mu_{smr}$ & $am$ & $\mu_{smr}$\\
    \hline 
    & Wilson & 1.038 & 1.12 & 0.4946 &  & 0.2929 & \\
    3.5 & Imp. Bri. & 0.6675 & 1.12 & 0.4517 & 0.56 & 0.316 & 0.4\\
    & DW & 0.728 & 0.38 & 0.495 &  & 0.3446 & \\
    \hline 
    & Wilson & 0.69 & 1.0 & 0.35 &  & 0.2 & \\
    3.0 & Imp. Bri. & 0.55 &  1.0 & 0.36 & 0.5 & 0.25 & 0.37\\
    & DW & 0.6 & 0.7 & 0.398 &  & 0.2785 & \\
    \hline 
    & Wilson & 0.45 &  & 0.2102 &  & 0.1105 & \\
    2.5 & Imp. Bri. & 0.416 &  0.84 & 0.268 & 0.45 & 0.184 & 0.3\\
    & DW & 0.465 &  & 0.303 &  & 0.2115 & \tabularnewline
    \hline 
    & Wilson & 0.2125 &  & 0.1 &  & 0.0267 & \\
    2.0 & Imp. Bri. & 0.2808 &  0.8 & 0.1705 & 0.4 & 0.119 & 0.25\\
    & DW & 0.3305 &  & 0.2154 &  & 0.149 & \\
    \hline 
    & Wilson & 0.0361 &  & $-$0.0197 &  & $-$0.061 & \\
    1.5 & Imp. Bri. & 0.1428 &  0.7 & 0.0789 & 0.35 & 0.059 & 0.15\\
    & DW & 0.191 &  & 0.1264 &  & 0.0765 & \\
    \hline 
    & Wilson & $-$0.1554 &  & $-$0.1447 &  & $-$0.1180 & \\
    1.0 & Imp. Bri. & 0.0182 & 0.63 & $-$0.0157 & 0.3 & $-$0.0036 & 0.12\\
    & DW & 0.0555 &  & 0.0408 &  & 0.012 & \\
    \hline 
  \end{tabular}
  \caption{
    Mass parameters given as inputs for each calculation.
    $m_{PS}$ is a target heavy-heavy (pseudo-scalar) meson mass.
    The bare mass parameter $am$ is listed for each fermion
    formulation: Wilson fermions, improved Brillouin fermions
    (Imp. Bri.), and smeared domain-wall fermions (DW). For unsmeared domain-wall fermions, see Appendix~\ref{sec:param_unsmr}.
    The gauge links are smeared in these measurements as described in the text. 
    $\mu_{smr}$ stands for a parameter appearing in the exponential function $e^{-\mu_{smr} r}$ to define the source.
    Since the critical mass is not subtracted, the bare mass for Wilson fermions (and for Imp. Bri.) can be negative.
  }
  \label{tab:param}
\end{table}

\subsection{Speed-of-light for pseudo-scalar meson}
The effective speed-of-light $c_{\mathrm{eff}}$ can be defined as
\begin{equation}
  c_{\mathrm{eff}}^2(\bm{p}) =
  \frac{E^2(\bm{p})-E^2(\bm{0})}{\bm{p}^2},
  \label{eq:speed-of-light}
\end{equation}
which is unity in the continuum theory and therefore gives a useful measure of the violation of Lorentz symmetry.

\begin{figure}[tbp]
  \centering
  \includegraphics[scale=0.3]{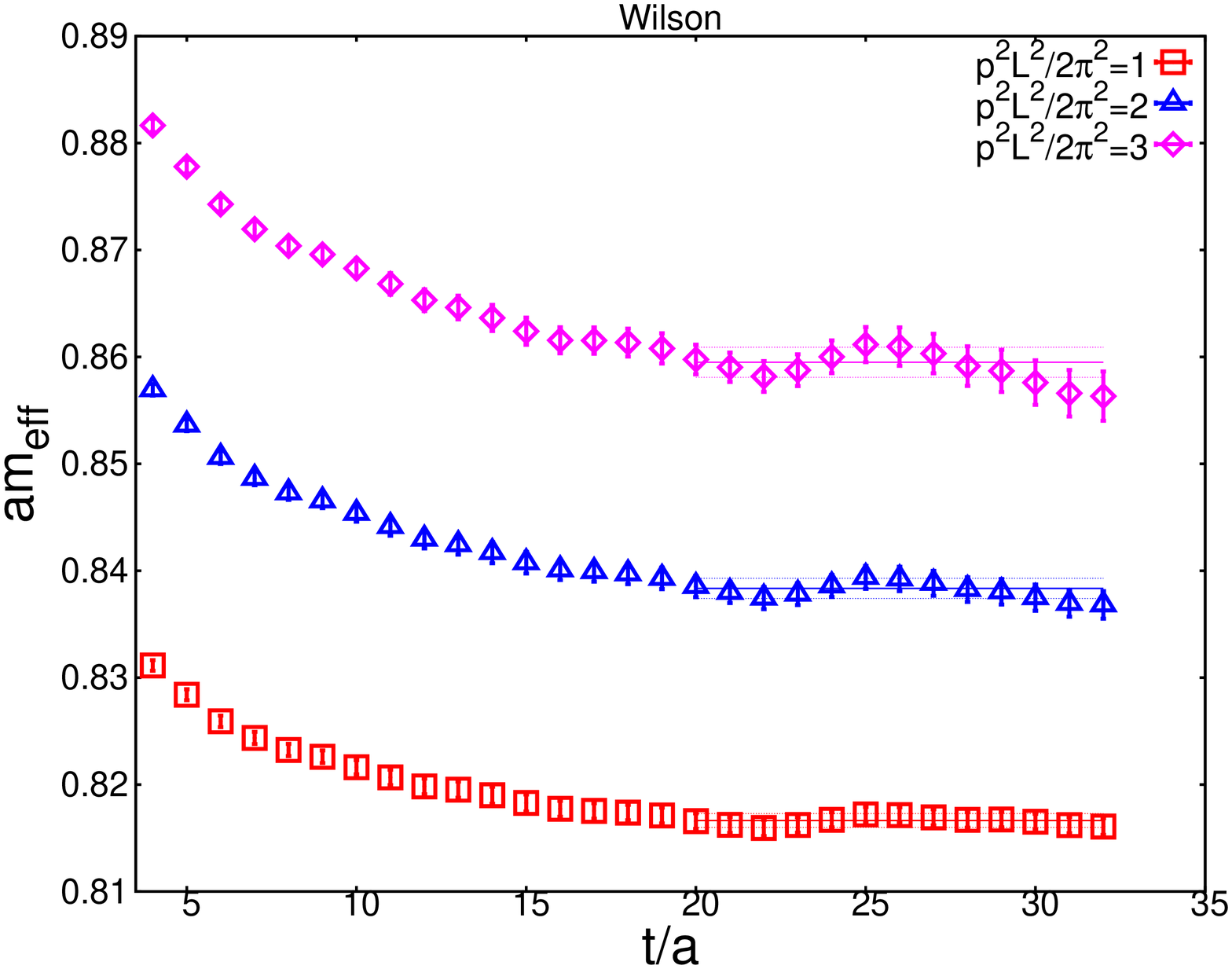}
  \includegraphics[scale=0.3]{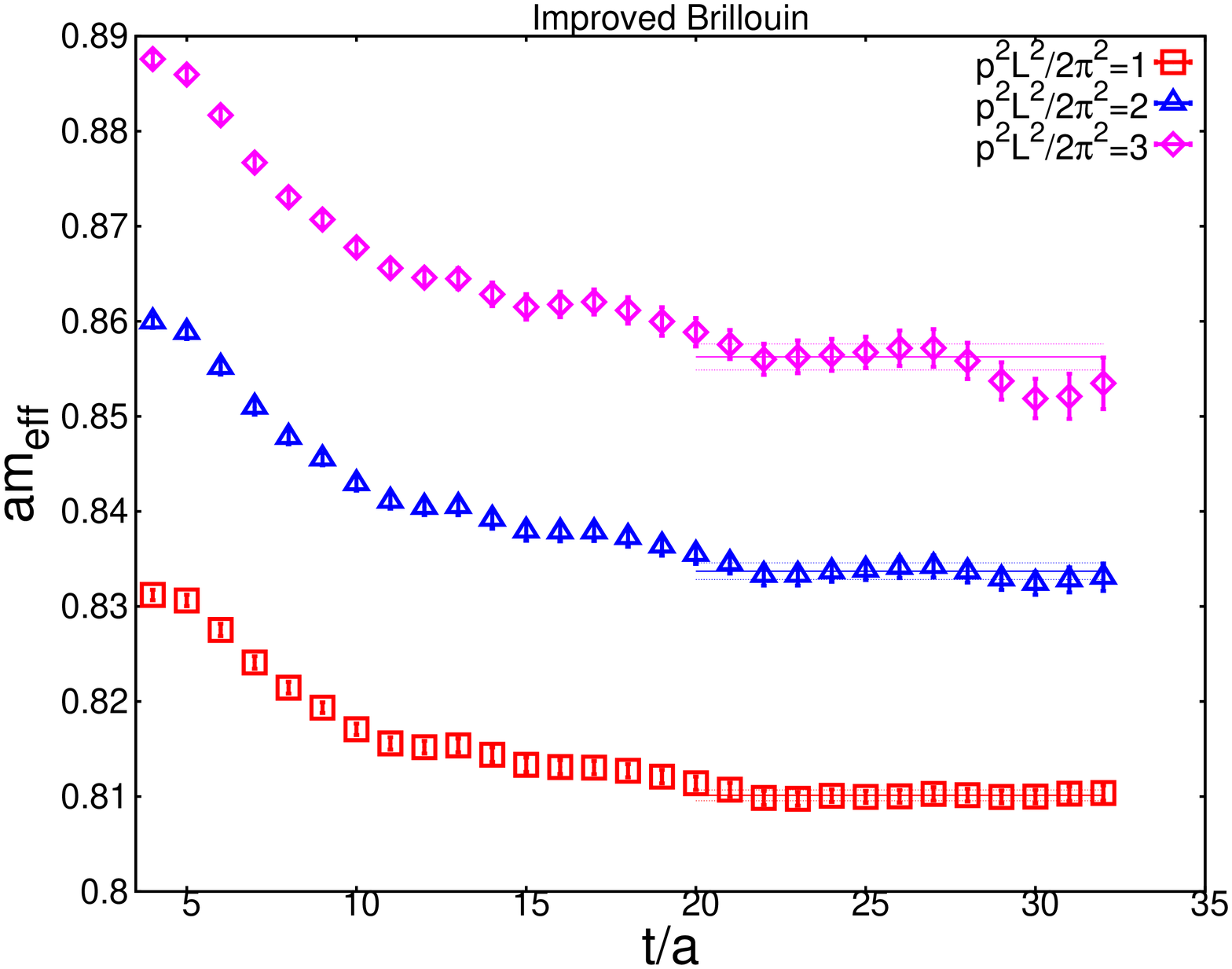}
  \includegraphics[scale=0.3]{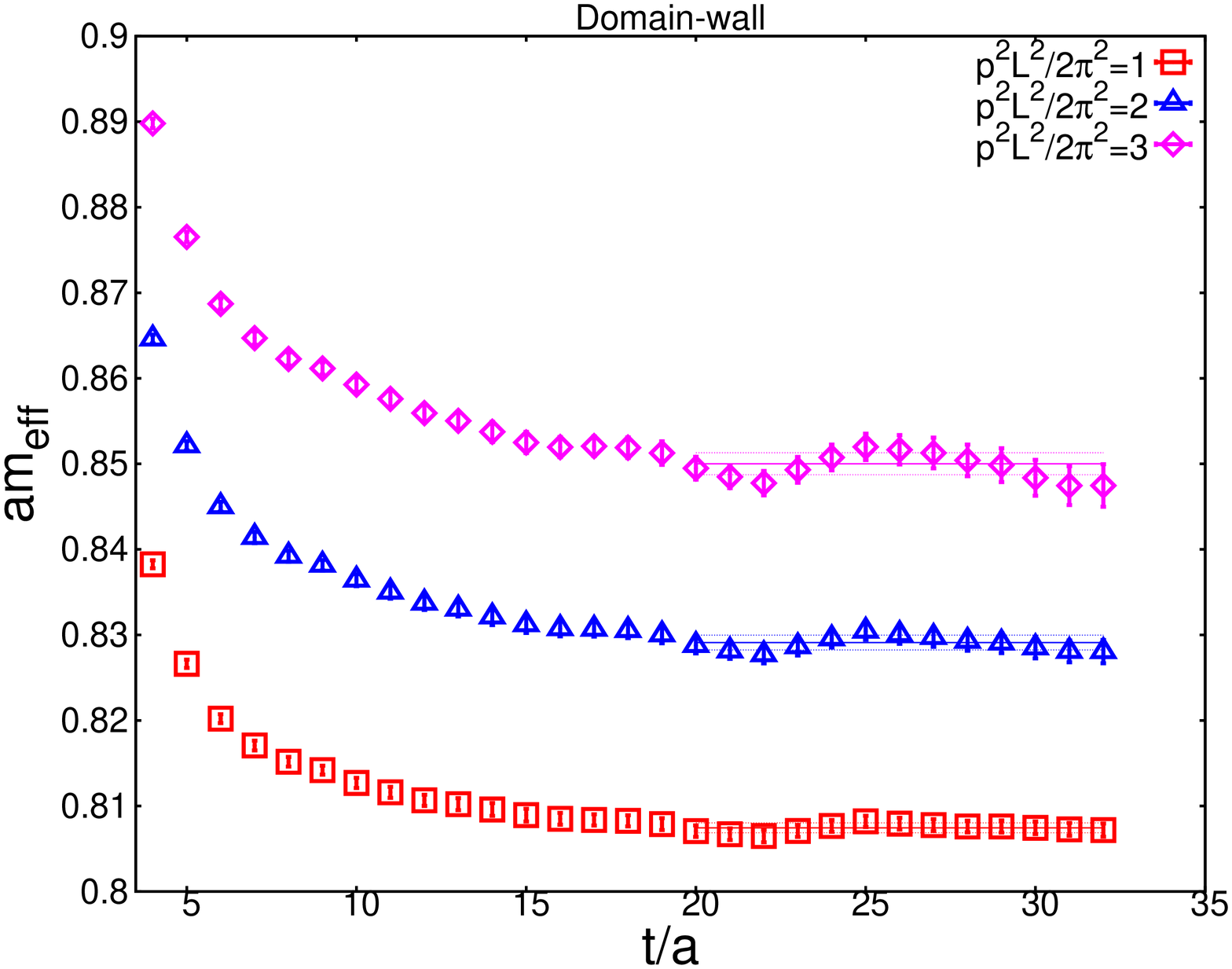}
  \caption{
    Effective mass of the pseudo-scalar meson with $\bm{p}^2/(2\pi/L)^2$ 
    = 1 (squares), 2 (triangles) and 3 (diamonds).
    Data corresponding to $m_{PS}$ = 3.0~GeV with Wilson fermions
    (top), improved Brillouin fermions (middle) and smeared domain-wall fermions (bottom) at $\beta$ = 4.89 are plotted. 
    Lines show the fit range and fitted value.
  }
  \label{fig:effm_typ_mom}
\end{figure}

We calculate the energy with lattice momenta $a\bvec{p}$ of (1,0,0), (1,1,0) and (1,1,1) in units of $2\pi/L$, where $L$ is almost the same for our ensembles. Effective masses for these finite momentum correlators are shown in Figure~\ref{fig:effm_typ_mom}.

As mentioned previously, the systematic error from the mass interpolation needs to be taken into account for the case of the unsmeared
domain-wall fermion data.  We interpolate the lattice data with various ans\"atze and take the spread of the central values as the systematic error. We find that this systematic error is subleading to the statistical error in all cases, but is particularly large for the case of the coarsest ensemble. The reason for this lies in the fact that we had to slightly extrapolate to reach $m_\mathrm{PS}$ = 3.0~GeV, causing the systematic error to be larger than on the other ensembles. This systematic error is added in quadrature to the statistical error before performing the continuum limit.

\begin{figure}[tbp]
  \includegraphics[scale=0.33]{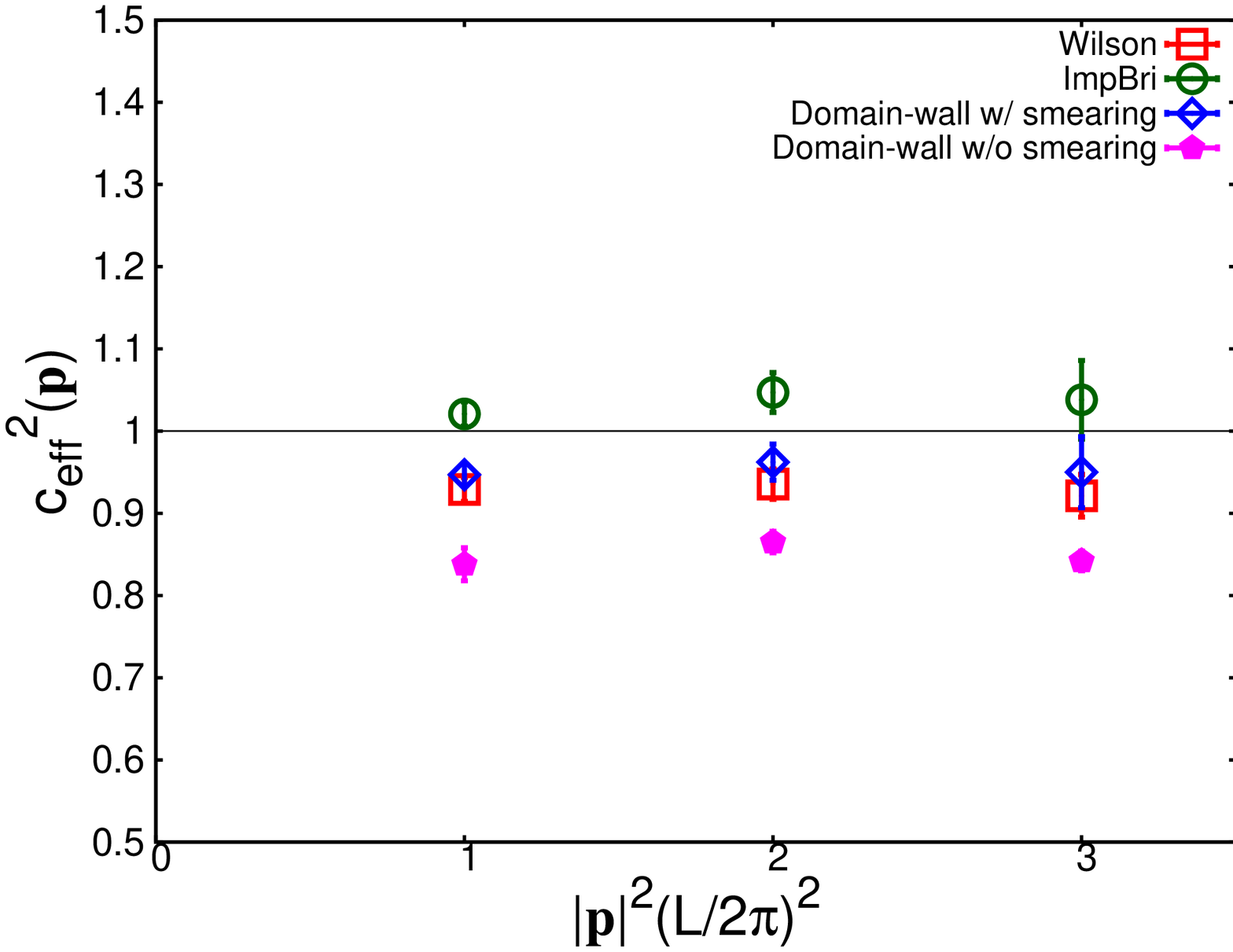}
  \includegraphics[scale=0.33]{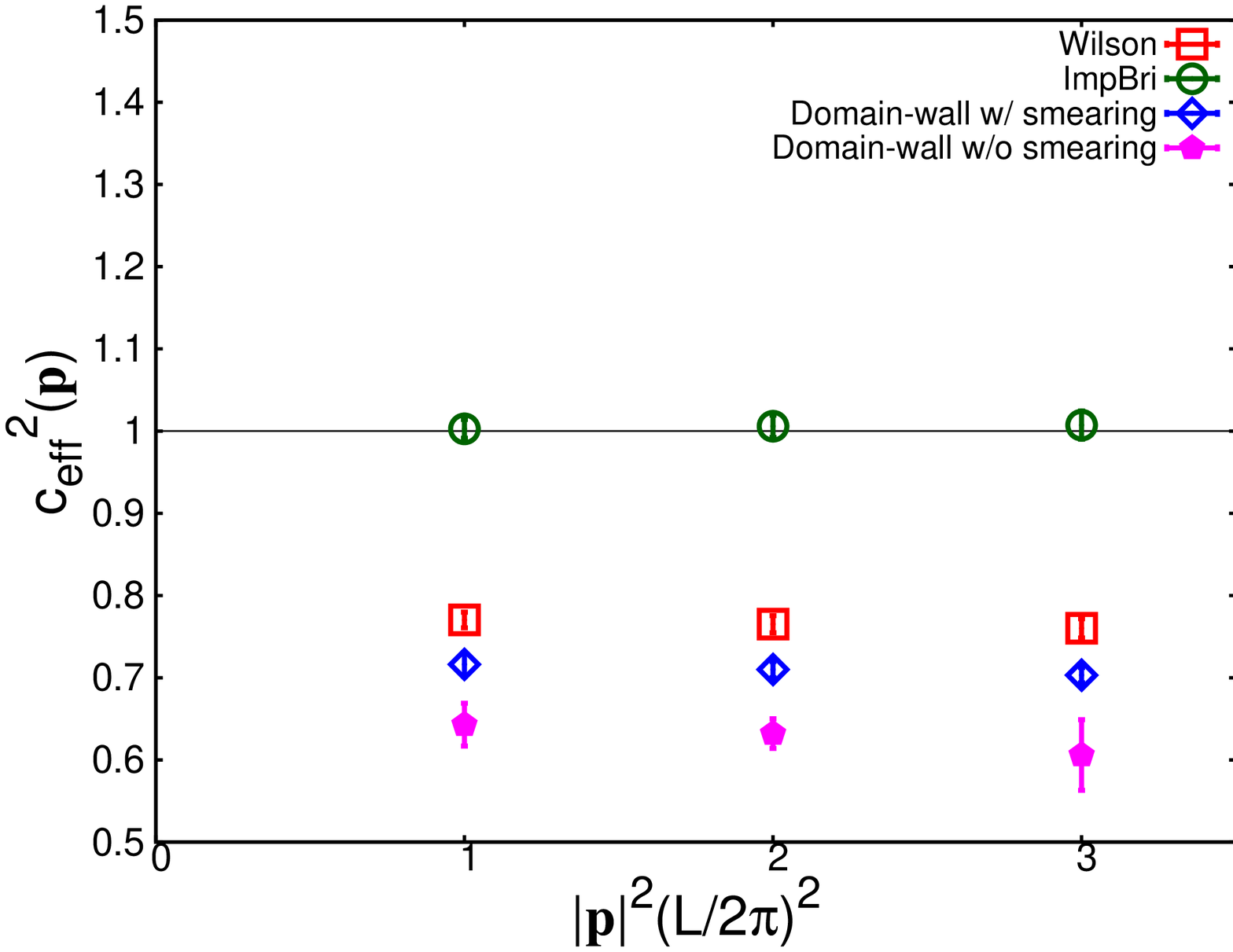}
  \caption{
    Effective speed-of-light calculated on the coarsest lattice
    ($\beta$ = 4.41) as a function of momentum squared.
    Data obtained with Wilson fermions (red squares), improved
    Brillouin fermions (green circles), smeared domain-wall fermions
    (blue diamonds), and unsmeared domain-wall fermions (filled magenta pentagons)
    are shown for pseudo-scalar meson masses at 1.5~GeV (left) and 3.0~GeV (right)
  }
  \label{fig:spl_v16}
\end{figure}

The effective speed-of-light thus calculated is plotted as a function of $|\bm{p}|^2/(2\pi/L)^2$ in Figure~\ref{fig:spl_v16} for the
heavy-heavy pseudo-scalar mesons of mass $m_{PS}$ = 1.5~GeV (left) and 3.0~GeV (right). The results on the coarsest lattice (at $\beta$ = 4.41) show substantial deviation from the continuum relation $c_{\mathrm{eff}}^2(\bm{p})=1$ for Wilson and domain-wall fermions. These two formulations are close to each other as far as the energy-momentum dispersion relation is concerned as the tree-level analysis suggests. For the lighter meson of $m_{PS}$ = 1.5~GeV, the deviation is already significant 5--10\%; for 3.0~GeV, which is close to charmonium, it is greater than 20\%.  The data for the improved Brillouin fermion are consistent with unity for both 1.5 and 3.0~GeV mesons. We calculated at four other heavy meson masses as listed in Table~\ref{tab:param}, and found that the results with the improved Brillouin fermion are always consistent with unity within the error.

\begin{figure}[tbp]
  \centering
  \includegraphics[scale=0.33]{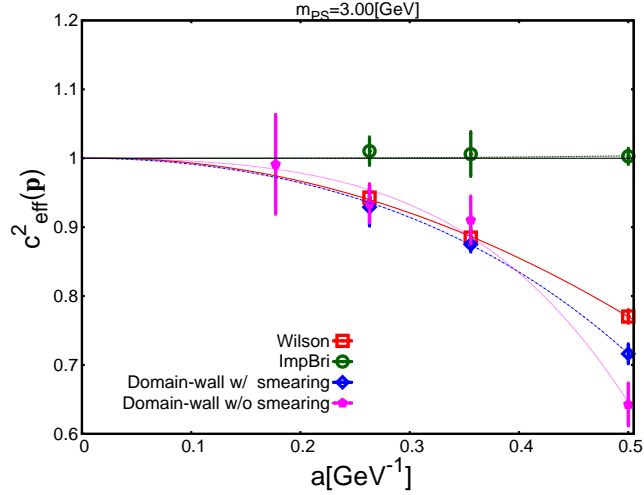}
  \caption{
    Scaling of the speed-of-light against $a$ 
    for the heavy-heavy meson mass $m_{PS}$ = 3.0~GeV.
    The results with $|\bm{p}|L/2\pi=1$ are shown.
    The different symbols are those of Wilson fermions (red squares),
    improved Brillouin fermions (green circles), smeared
    domain-wall fermions (blue diamonds), and unsmeared domain-wall
    fermions (filled magenta pentagons).
    For the details on the fit curves, see the text.
  }
  \label{fig:scaling_spl}
\end{figure}

We show the scaling of the speed-of-light in Figure~\ref{fig:scaling_spl} against the lattice spacing $a$. The data for momentum $|\bm{p}|L/2\pi=1$ are shown; higher momenta are similar but have larger error bars. As one can see, for the heavy-heavy meson of mass $m_{PS}$ = 3.0~GeV, no deviation from the continuum relation $c_{\mathrm{eff}}^2(\bm{p})=1$ is found with the improved Brillouin fermion in the range of lattice spacing $a\lesssim$ 0.1~fm. The results with the Wilson and domain-wall fermions are similar. A substantial deviation of 20--30\% is found on the coarsest lattice, which decreases to the level of 5\% at $a\simeq$ 0.05~fm.  This would be a typical size of discretization error for these lattice formulations, unless other theoretical constrains such as that of non-relativistic effective theory \cite{ElKhadra:1996mp} are introduced. 

In order to quantify the size of scaling violations, we attempt to model the discretization effect using the data at $|\bvec{p}|=2\pi/L$. Assuming that the continuum relation $c_{\mathrm{eff}}^2(\bm{p})=1$ is recovered in the continuum limit, we employ an ans\"atz $f(a)=1+c_1a+c_2a^2$ for Wilson fermions. For smeared and unsmeared domain-wall fermions, an ans\"atz $f(a)=1+c_2a^2+c_4a^4$ is used instead, because the $O(a)$ and $O(a^3)$ terms are forbidden by chiral symmetry. 
Strictly speaking, there might be $O(a)$ and $O(a^3)$ discretization effects because of non-zero $m_{res}$, but we assume that the residual breaking of chiral symmetry is negligible in our setup. For the improved Brillouin fermion, the leading discretization effects are those of $O(a^3)$. We therefore assume a function $f(a)=1+c_3a^3$.

For each fermion formalism, we obtain a reasonable quality of fit with $\chi^2/$d.o.f $\lesssim 0.5$. The fit results are $c_1$ = 0.03(11)~GeV and $c_2$ = $-$1.0(2)~GeV$^2$ for Wilson fermions, $c_2$ = $-$0.9(2)~GeV$^2$ and $c_4$ = $-$1.1(8)~GeV$^4$ for smeared domain-wall fermions.  For the unsmeared one, we obtain $c_2$ = $-$0.4(4)~GeV$^2$ and $c_4$ = $-$4.2(1.8)~GeV$^4$. Also, $c_3$ = 0.03(9)~GeV$^3$ for improved Brillouin fermions at  $|{\bf p}|=2\pi/L$ is obtained. These fit results are plotted in Figure~\ref{fig:scaling_spl}. These results suggest that the coefficients have a reasonable size of $O(\mbox{1~GeV})$ or less. The coefficient for the improved Brillouin fermion is essentially zero even at the order of $a^3$.

Since improved Brillouin fermions are designed to achieve $O(a)$ and $O(a^2)$ improvement only in the free theory, there is a possibility that significant contributions of $O(a)$ and $O(a^2)$ appear due to radiative corrections. A naive order-counting suggests that their size is $O(\alpha_s a)$ or $O(\alpha_s a^2)$, respectively. Assuming $\alpha_s\sim$ 0.2--0.3, these contributions are not negligible. The small scaling violation of the actual data may suggest that these effects are small, which is consistent with an expectation that the radiative corrections are relatively small in general when using link smearing \cite{DeGrand:2002va}. An explicit perturbative calculation to confirm this expectation is on-going. 

\subsection{Hyperfine splitting}
The hyperfine splitting $m_{V}-m_{PS}$ is also an interesting quantity to investigate scaling violations. In the non-relativistic effective theory, it arises from the Pauli term of the form $\psi^\dagger\bm{\sigma}\cdot\bm{B}\psi$. As this term has the same form as the clover term of the $O(a)$-improved action, it is expected that the hyperfine splitting is very sensitive to the $O(a)$ discretization effects and also possibly to higher order effects.

\begin{figure}[tbp]
  \centering
  \includegraphics[scale=0.33]{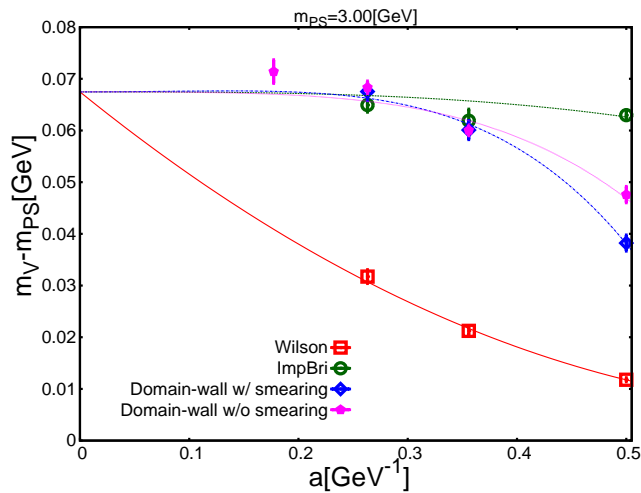}
  \caption{
    Scaling of the hyperfine splitting against $a$ for the heavy-heavy
    meson mass $m_{PS}$ = 3.0~GeV. 
    The different symbols are those of the Wilson fermion (red squares),
    the improved Brillouin fermion (green circles), the smeared
    domain-wall fermion (blue diamonds) and the unsmeared domain-wall
    fermion (filled magenta pentagons). 
    For the details on the fit curves, see the text.
  }
  \label{fig:scaling_hyp_c3}
\end{figure}

We show the scaling of $m_V-m_{PS}$ in Figure~\ref{fig:scaling_hyp_c3} at $m_{PS}=3.0$ GeV. For this quantity, the value in the continuum limit is not known. Experimentally, the charmonium hyperfine splitting is 117~MeV, but in the quenched theory it could be significantly different from this value. We therefore do not assume that the continuum limit of the lattice data will reproduce the experimental value. The result in Figure~\ref{fig:scaling_hyp_c3} clearly shows substantial scaling violations for the Wilson fermion. The splitting is several times smaller than the results from other formulations. Particularly on the coarsest lattice this can be seen ($a\simeq$ 0.1~fm). This is in accordance with the expectation that the hyperfine splitting is sensitive to $O(a)$ effects. With domain-wall fermions (both smeared and unsmeared), we also find a significant discretization effect, though much less severe than in the case of Wilson fermions. In contrast, the result with improved Brillouin fermions shows very mild $a$ dependence. From their results alone, one cannot tell any sign of the discretization effects.

\begin{figure}[tbp]
  \centering
  \includegraphics[scale=0.33]{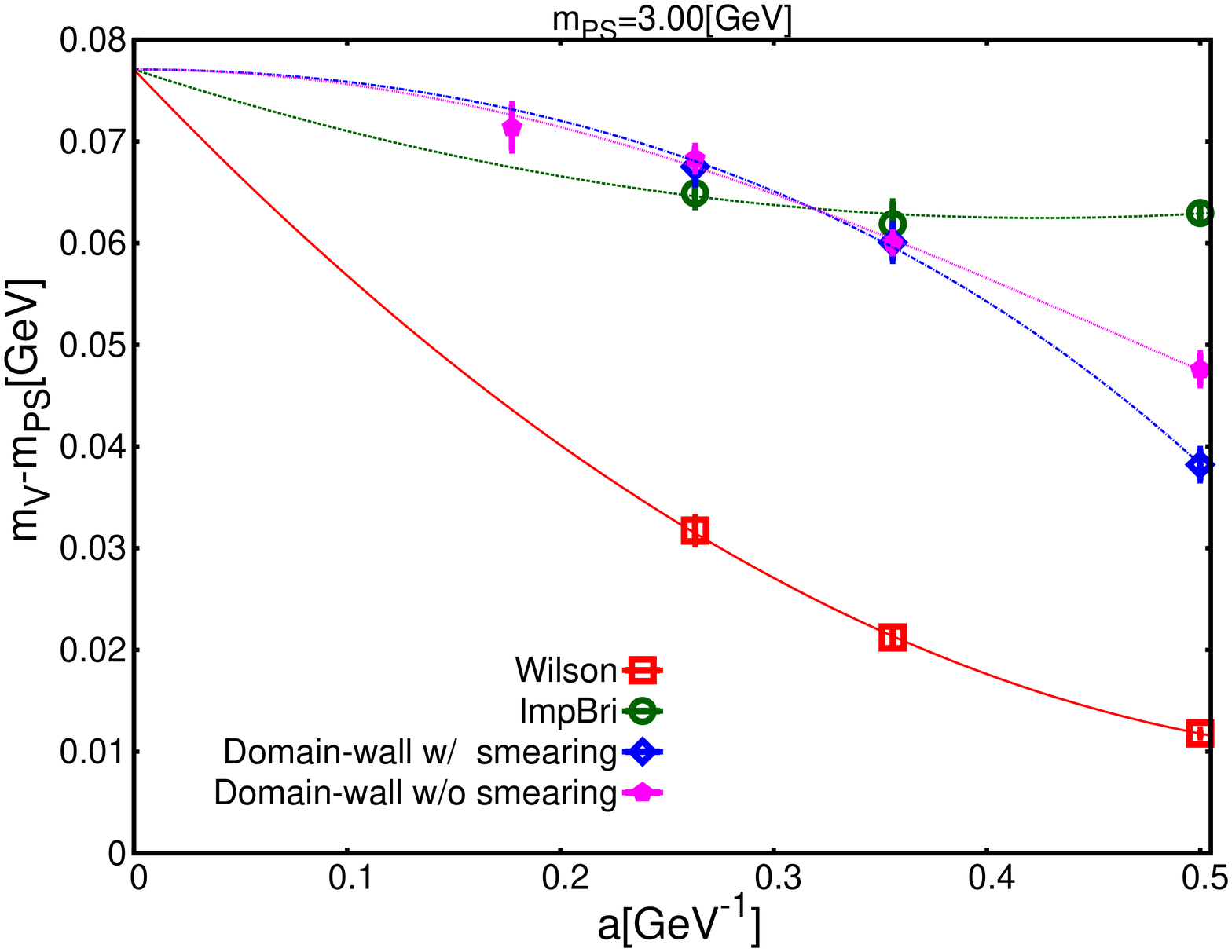}
  \caption{
    Same as Figure~\ref{fig:scaling_hyp_c3}, but with a fit function
    $f(a)=1+c_1a+c_2a^2$ for the improved Brillouin fermion. 
  }
  \label{fig:scaling_hyp}
\end{figure}

Here again, we examine the scaling violation by fitting the lattice data as a function of $a$. Since the value in the continuum limit is unknown, we assume that four lattice formulation give the universal result in the continuum limit and fit all the data simultaneously. For Wilson fermions we employ $f(a)=c_0+c_1a+c_2a^2$, while for smeared and unsmeared the domain-wall fermions we take $f(a)=c_0+c_2a^2+c_4a^4$ as constrained by chiral symmetry. For improved Brillouin fermions, we first attempt a fit with a function $f(a)=c_0+c_3a^3$. As shown in Figure~\ref{fig:scaling_hyp_c3}, a combined fit of four formulations is unsuccessful ($\chi^2/d.o.f.\simeq$ 3.8). 
The continuum limit of this fit yields $c_0$ = 0.068(1)~GeV. 
 
It may indicate that the improved Brillouin action receives significant radiative correction at $O(a)$ (and $O(a^2)$). 
Therefore we also try to fit with $f(a)=c_0+c_1a+c_2a^2$ for the action as that for Wilson fermions. The quality of the fit is better ($\chi^2/d.o.f.\simeq 0.3$) and the central value for $c_0$ is slightly higher: $c_0$ = 0.077(3)~GeV. 
The results are shown in Figure~\ref{fig:scaling_hyp}. The fit that allows discretization effects of $O(a)$ and $O(a^2)$ indicates that the coefficients for improved Brillouin fermions ($c_1$ = $-$0.07(2)~GeV$^2$, $c_2$ = 0.08(3)~GeV$^3$) are much smaller than those of  Wilson fermions ($c_1$ = $-$0.22(2)~GeV$^2$, $c_2$ = 0.18(2)~GeV$^3$). One has to be careful about the result for Brillouin fermions, because if this fit captures the actual discretization effect there is a significant cancellation between the $O(a)$ and $O(a^2)$ terms and the data points between $a$ = 0.25 and 0.5~GeV$^{-1}$ show a flat behavior. More detailed scaling analysis of other quantities need to be performed if it is the case.
For smeared and unsmeared domain-wall fermions, $c_2$ = $-$0.12(4)~GeV$^3$, $c_4$ = $-$0.14(15)~GeV$^5$ and $c_2$ = $-$0.15(4)~GeV$^3$, $c_4$ = $-$0.11(13)~GeV$^5$ are obtained, respectively. The size of discretization effects for these formulations is very similar, though the data point of smeared domain-wall fermion at $a$ = 0.5~GeV$^{-1}$ shows significantly larger deviation from the continuum limit. 
Also we observe that our value of the continuum limit is consistent with that of Ref.  \cite{Choe:2003wx} in which the charmonium spectra are simulated on quenched lattices with some different valence quark formalisms.

\begin{figure}[tbp]
  \centering
  \includegraphics[clip,scale=0.40]{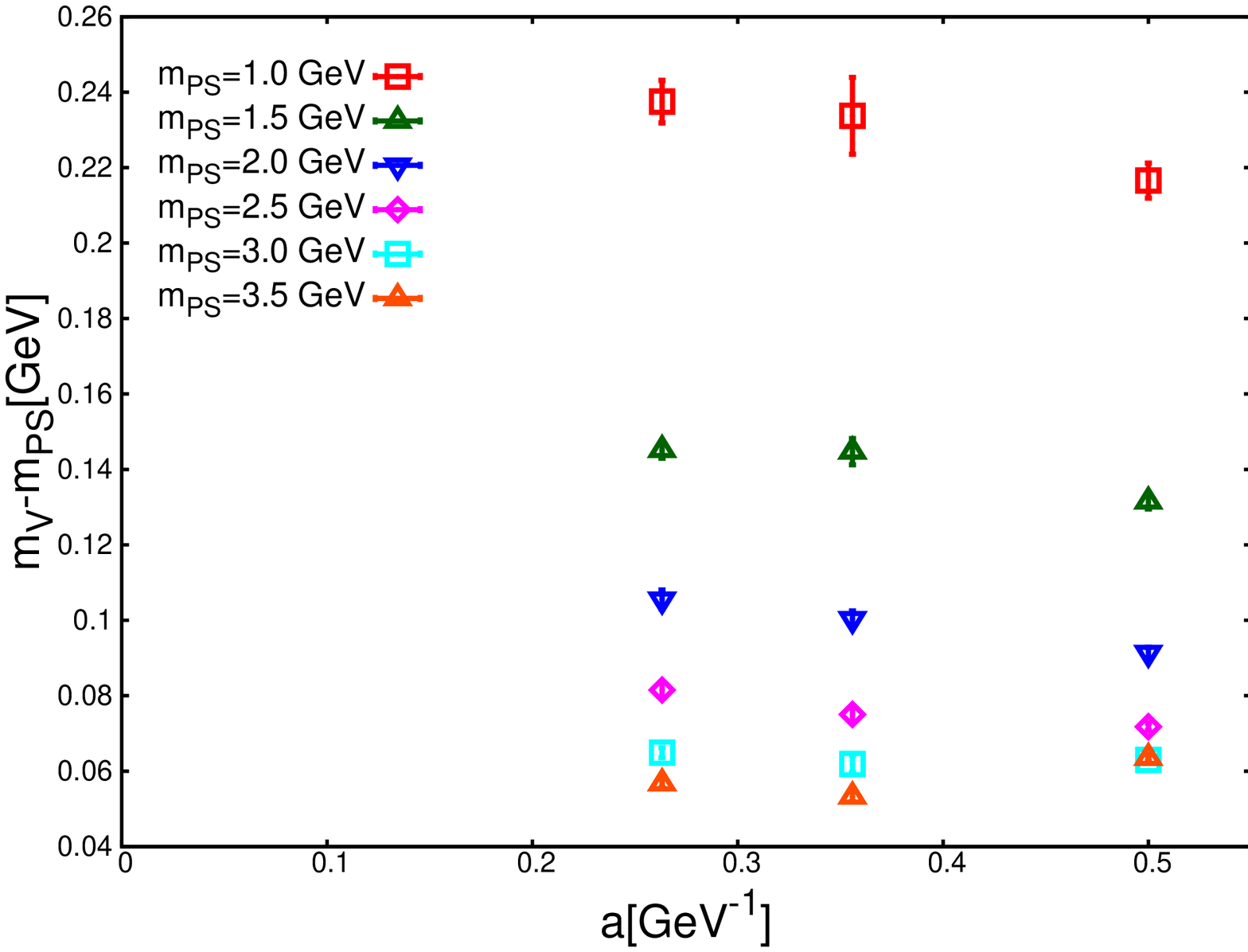}\\
  \includegraphics[clip,scale=0.40]{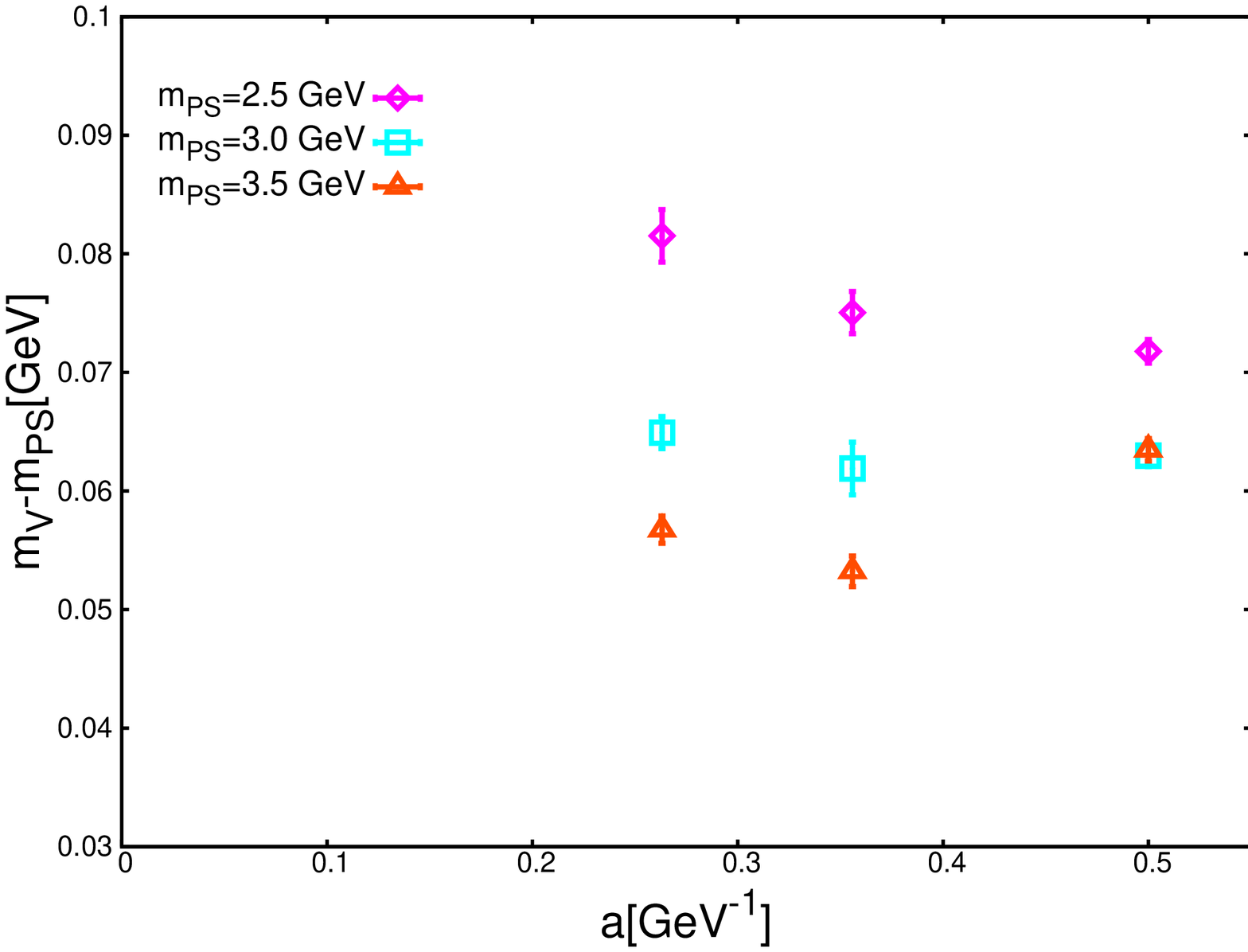}
  \caption{
    Hyperfine splitting of the heavy-heavy mesons of mass
    $m_{PS}$ = 1.0, 1.5, 2.0, 2.5, 3.0, and 3.5~GeV 
    calculated with the improved Brillouin fermion action
    (top panel).
    The lattice results with the improved Brillouin fermion are
    plotted as a function of lattice spacing $a$.
    The plot in the bottom panel enlarges the results for three
    heaviest masses. 
  }
  \label{fig:hyp_ibr_scaling}
\end{figure}

Finally, we study the scaling of the hyperfine splitting for different sets of heavy quark masses, in order to check the discretization effects for the improved Brillouin fermion action. Figure~\ref{fig:hyp_ibr_scaling} shows the hyperfine splitting for various heavy-heavy meson masses plotted against the lattice spacing $a$. A good scaling is observed in general, but focusing on the heaviest mass ($m_{PS}$ = 3.5~GeV) we observe a very strong scaling violation for the coarsest lattice point $a\simeq$ 0.5~GeV$^{-1}$. At this lattice spacing, the natural heavy quark scaling that the hyperfine splitting decreases for heavier masses is lost between $m_{PS}$ = 3.0~GeV and 3.5~GeV. This may indicate the problem of the improved action for too large $am$ as discussed in Section~\ref{sec:heavy_problem}.  The bare mass for these two masses is 0.55 and 0.67, respectively, which is in the mass range where the effects of doublers could become significant. 

\section{Conclusion}
\label{sec:conclusion}
The energy-momentum dispersion relation is a key property of relativistic field theories.  On the lattice, the Lorentz symmetry is explicitly broken by the lattice discretization and the continuum dispersion relation is expected to be recovered only in the continuum limit.  
For practical applications, how fast one can approach the continuum limit becomes a crucial question;  while charm quarks can be simulated with moderately large values of the input quark mass in lattice units $am$ on ensembles available today, the bottom quark mass cannot be made much less than 0.5--1.0.
It is therefore important to design a lattice formulation that respects the symmetry relation of the continuum theory as much as possible. The Brillouin fermion is among such class of fermion formulations, and in this paper we consider its further improvement and carry out the corresponding numerical tests.

The improved Brillouin fermion defined by (\ref{eq:imp_bri_action}) has the energy-momentum dispersion relation which is close to the continuum one. This is confirmed at the tree-level for the massless case ($am=0$) and the massive case ($am=0.5$). The leading discretization effect is $O(a^3)$, but as far as the dispersion relation is concerned, the actual error seems to be much smaller than a naive estimate $O((am)^4)\sim$ 13\% for $am=0.5$. Through the radiative correction, the terms of $O(a\alpha_s)$ and $O(a^2\alpha_s)$ are induced, and their effects have to be carefully examined. In our non-perturbative test in the quenched theory, the discretization effect is insignificant on the lattices of $1/a$ = 2.0--5.6~GeV and charm quark mass $m\simeq$ 1.3~GeV. The sign of cancelling $O(a)$ and $O(a^2)$ effects found in the hyperfine splitting needs to be studied more carefully, though. The most direct approach would be to calculate on finer lattices. We also to plan to inspect the size of corresponding one-loop correction. In successful, the action is a promising candidate for the simulation of charm quark on more realistic unquenched lattices.

For more extensive calculations, the numerical cost would become an important issue. With our current implementation, the inversion of charm quark propagators takes 2--3 times more time than for the inversion of the domain-wall fermion. It would be worth spending such numerical cost given the highly suppressed discretization effects, but further improvement of the numerical code is certainly desired.

Another important conclusion from our analysis is that the continuum extrapolation with the (smeared and unsmeared) domain-wall fermion is possible for charm quark, provided that the data are available in the region beyond $1/a\gtrsim$ 3.0~GeV. A combined fit including other formulation as done in this work would be useful to have better control of systematic effects.

\vspace*{0.8cm}

We thank Stephan D\"{u}rr for detailed discussions and for disclosing a sample code for the Brillouin operator.
Numerical simulations are performed on the IBM System Blue Gene Solution at High Energy
Accelerator Research Organization (KEK) under a support of its Large Scale Simulation Program
(No. 13/14-04). This work is supported in part by the Grant-in-Aid of the Japanese Ministry of
Education (No. 26247043, 26400259) and the SPIRE (Strategic Program for Innovative Research) Field5 project.
The research leading to these results has received funding from the European Research Council under the European Union's Seventh Framework Programme (FP7/2007-2013) / ERC Grant agreement 279757.
The authors acknowledge the use of the IRIDIS High Performance Computing Facility, and associated support services at the University of Southampton, in the completion of this work. 
The program package GLU used for the Wilson flow measurements was kindly provided by Jamie Hudspith (https://github.com/RJhudspith/GLU). The authors gratefully acknowledge computing time granted through the STFC funded DiRAC facility (grants ST/K005790/1, ST/K005804/1, ST/K000411/1, ST/H008845/1)

\bibliographystyle{JHEP}
\bibliography{draft}

\newpage
\appendix
\section{Implementation of gauge invariant operators}
\label{sec:hops}
In order to keep the rotational symmetry under the cubic group, one
has to average over possible paths connecting the interacting points.
The most economical way is to take the paths of minimum taxi-driver
distance.
The average can be represented as off-axis link variables.
For instance, for the interaction in a $\hat{\mu}$-$\hat{\nu}$ plane
connected by two links we may define
\begin{eqnarray}
  V_{\hat{\mu}+\hat{\nu}}(n) 
  & = & 
        \frac{1}{2}\left[
        U_\mu(n)U_\nu(n+\hat{\mu}) +
        U_\nu(n)U_\mu(n+\hat{\nu})
        \right],
        \nonumber\\
  V_{-\hat{\mu}-\hat{\nu}}(n)
  & = &
        \frac{1}{2}\left[
        U_\mu^\dagger(n-\hat{\mu})U_\nu^\dagger(n-\hat{\mu}-\hat{\nu})+
        U_\nu^\dagger(n-\hat{\nu})U_\mu^\dagger(n-\hat{\mu}-\hat{\nu})
        \right],
        \nonumber\\
  V_{\hat{\mu}-\hat{\nu}}(n)
  & = &
        \frac{1}{2}\left[
        U_\mu(n)U_\nu^\dagger(n+\hat{\mu}-\hat{\nu}) + 
        U_\nu^\dagger(n-\hat{\nu})U_\mu(n-\hat{\nu}
        )\right],
        \nonumber\\
  V_{-\hat{\mu}+\hat{\nu}}(n)
  & = &
        \frac{1}{2}\left[
        U_\nu(n)U_\mu^\dagger(n+\hat{\nu}-\hat{\mu}) +
        U_\mu^\dagger(n-\hat{\mu})U_\nu(n-\hat{\mu})
        \right].
        \nonumber\\
\end{eqnarray}
For 3-hop and 4-hop terms, there are off-axis link variables like
\begin{eqnarray}
  V_{\hat{\mu}+\hat{\nu}+\hat{\rho}}\left(n\right) 
  & = &
        \frac{1}{3}\left[
        V_{\hat{\mu}+\hat{\nu}}(n)U_\rho(n+\hat{\mu}+\hat{\nu}) + 
        V_{\hat{\mu}+\hat{\rho}}(n)U_\nu(n+\hat{\mu}+\hat{\rho}) 
        \right.
        \nonumber\\
  & &
        \left. + 
        V_{\hat{\nu}+\hat{\rho}}(n)U_\mu(n+\hat{\nu}+\hat{\rho})
        \right],
        \nonumber\\
  V_{\hat{\mu}+\hat{\nu}+\hat{\rho}+\hat{\sigma}}(n) 
  & = & 
        \frac{1}{4}\left[
        V_{\hat{\mu}+\hat{\nu}+\hat{\rho}}(n)
        U_\sigma(n+\hat{\mu}+\hat{\nu}+\hat{\rho})
        \right.
        \nonumber\\
  & &
        \left. +
        V_{\hat{\mu}+\hat{\nu}+\hat{\sigma}}(n)
        U_{\rho}(n+\hat{\mu}+\hat{\nu}+\hat{\sigma})
        \right.
        \nonumber\\
  & &
      \left. + 
      V_{\hat{\mu}+\hat{\rho}+\hat{\sigma}}(n)
      U_{\nu}(n+\hat{\mu}+\hat{\rho}+\hat{\sigma})
      \right.
      \nonumber\\
  & &
      \left. +
      V_{\hat{\nu}+\hat{\rho}+\hat{\sigma}}(n)
      U_{\mu}(n+\hat{\nu}+\hat{\rho}+\hat{\sigma})
      \right].
\end{eqnarray}
Note that the Brillouin fermion has 80 nearest-neighbors within a
$3^4$ hypercube, and thus 80 off-axis link variables have to be
prepared. 
Calculation of these off-axis links can be done before the conjugate
gradient iterations start, as the link variable is unchanged during
the solver steps.
This method is called ``overall smearing strategy (OSS)''
\cite{Durr:2010ch}. 


We also consider an alternative method to calculate the covariant
Brillouin Laplacian and the isotropic derivative operators with gauge
fields. 
We use the following recursive definition.
\begin{eqnarray}
  a\triangle^{bri}(n,m) \psi_{m} 
  & = &
        \frac{1}{64} \sum_\mu D_\mu^+\psi_n^{\prime\prime\prime} -
        \frac{15}{4} \psi_{n},
  \nonumber\\
  \psi_n^{\prime\prime\prime} 
  & \equiv & 
             8\psi_n + 
             \frac{1}{2}\sum_{\nu\neq\mu} D_\nu^+ \psi_{n}^{\prime\prime},
  \nonumber\\
  \psi_n^{\prime\prime} 
  & \equiv &
             4\psi_{n} + \frac{1}{3}\sum_{\rho\neq\mu,\nu} D_\rho^+
             \psi_n^\prime,
  \nonumber\\
  \psi_n^\prime 
  & \equiv &
             2\psi_{n} +
             \frac{1}{4}\sum_{\sigma\neq\mu,\nu,\rho} D_\sigma^+ \psi_n,
\end{eqnarray}
where $D_{\mu}^{\pm}$ is defined by 
\begin{equation}
  D_\mu^\pm \psi_n =
  U_\mu(n)\psi_{n+\hat{\mu}} \pm
  U_\mu^\dagger(n-\hat{\mu})\psi_{n-\hat{\mu}}.
\end{equation}
We can write down a similar formula for the isotropic derivative.
That is
\begin{eqnarray}
  \nabla_x^{iso}(n,m)\psi_{m} 
  & = & 
        \frac{1}{432}\left(
        D_x^-\xi_n^{\prime\prime\prime} +
        \frac{1}{2}\sum_{\nu\neq x} D_\nu^+\eta_n^{\prime\prime\prime}
        \right),
        \nonumber\\
  \xi_n^{\prime\prime\prime} 
  & \equiv &
             64\psi_n
             + \frac{1}{2} \sum_{\nu\neq x} D_\nu^+
             \xi_n^{\prime\prime},
             \nonumber\\
  \xi_n^{\prime\prime} 
  & \equiv &
             16\psi_n
             + \frac{1}{3}\sum_{\rho\neq x,\nu} D_\rho^+ \xi_n^\prime,
             \nonumber\\
  \xi_n^\prime 
  & \equiv &
             4\psi_n
             + \frac{1}{4}\sum_{\sigma\neq x,\nu,\rho} 
             D_\sigma^+ \psi_n,
             \nonumber\\
  \eta_n^{\prime\prime\prime} 
  & \equiv &
             D_x^- \xi_n^{\prime\prime}
             + \frac{1}{3}\sum_{\rho\neq x,\nu} D_\rho^+
             \eta_n^{\prime\prime},
             \nonumber\\
  \eta_n^{\prime\prime} 
  & \equiv &
             D_x^-\xi_n^\prime
             + \frac{1}{4}\sum_{\sigma\neq x,\nu,\rho}
             D_\sigma^+ \eta_n^\prime,
             \nonumber\\
  \eta_n^\prime & \equiv & D_x^-\psi_n.
\end{eqnarray}
This recursive formula gives the same result as OSS does.
This compact form is also useful for perturbative calculation using
the automatic calculation package such as \cite{Lehner:2012bt}. 

Computational codes for both options are implemented in the Iroiro++
package \cite{Cossu:2013ola}.

\newpage
\section{Simulated parameters for the unsmeared M\"obius  domain-wall  fermion}
\label{sec:param_unsmr}

\begin{table}[!h]
  \centering
  \begin{tabular}{ccccc}
    \hline 
     $L/a$ &  & $am$ &  & smearing \\
     & start & step & end &  parameter\\
     \hline
    16 & 0.1 & 0.05 & 0.4 & 2.8  \\
    24 & 0.1 & 0.05 & 0.4 & 4.0  \\
    32 & 0.1 & 0.05 & 0.4 & 5.6 \\
    48 & 0.04 & 0.04 & 0.28 & 11.7 \\
    \hline
    16 & 0.1 & 0.05 & 0.4 & 4.5  \\
    24 & 0.066 & 0.033 & 0.396 & 6.0  \\
    32 & 0.07 & 0.04 & 0.39 & 8.8 \\
    48 & 0.04 & 0.04 & 0.28 & 11.7  \\
    \hline 
  \end{tabular}
  \caption{
    Simulated bare quark mass in lattice units for the unsmeared M\"obius domain-wall fermion. 
    The masses starting from ``start'' with a step of ``step'' and ending at ``end'' are simulated.
    The ``smearing parameter'' refers to the choice of the smearing parameter for the Gaussian smearing of the source/sink of the
    propagators. 
    The first block corresponds to the dispersion relation
    measurements with a point source, the second block to the
    hyperfine splitting measurements with $Z_2$-wall source. 
    All measurements we carried out with the unsmeared M\"obius domain-wall fermion
    are with parameters $L_s$ = 12 and $M_0$ = 1.6.
  }
  \label{tab:param_unsmr}
\end{table}

\end{document}